\documentclass[fleqn,usenatbib]{mnras}
\usepackage{newtxtext,newtxmath}
\usepackage[T1]{fontenc}
\DeclareRobustCommand{\VAN}[3]{#2}
\let\VANthebibliography\thebibliography
\def\thebibliography{\DeclareRobustCommand{\VAN}[3]{##3}\VANthebibliography}

\usepackage{graphicx}	
\usepackage{amsmath}	
\usepackage[para,online,flushleft]{threeparttable}
\usepackage{soul} 
\usepackage{subcaption}

\usepackage{graphicx}
\usepackage{float}

\usepackage{longtable}
\usepackage{lscape}      
\usepackage{multicol}    
\usepackage{booktabs}   

\usepackage{url}
\title[Semi-Supervised Models for Repeating FRBs]{Revealing Hidden Repeaters in the CHIME/FRB Catalog: Semi-Supervised Insights into the Fast Radio Burst Population}

\author[N. Mankatwit et al.]{
N. Mankatwit,$^{1}$
P. Thongkonsing,$^{1}$
S. Loekkesee,$^{2}$
P. Chainakun,$^{1,3}$\thanks{E-mail: pchainakun@g.sut.ac.th}
W. Luangtip,$^{2}$\thanks{E-mail: wasutep@g.swu.ac.th }
S. Sanpa-arsa$^{4}$\thanks{E-mail: s.tuck.sanpaarsa@gmail.com}
\\
$^{1}$School of Physics, Institute of Science, Suranaree University of Technology, 111 University Avenue, Muang, Nakhon Ratchasima, 30000, Thailand\\
$^{2}$Department of Physics, Faculty of Science, Srinakharinwirot University, Bangkok 10110, Thailand
\\
$^{3}$Centre of Excellence in High Energy Physics and Astrophysics, Suranaree University of Technology, Nakhon Ratchasima 30000, Thailand\\
$^{4}$National Astronomical Research Institute of Thailand (Public Organization), 260 Moo 4, Donkaew, Mae Rim, Chiang Mai, 50180, Thailand
}

\date{Accepted XXX. Received YYY; in original form ZZZ}

\pubyear{\the\year{}}

\begin{document}
\label{firstpage}
\pagerange{\pageref{firstpage}--\pageref{lastpage}}
\maketitle

\begin{abstract}

Fast radio bursts (FRBs) are millisecond-duration extragalactic transients, observationally classified as repeaters or non-repeaters. This classification may be biased, as some apparently non-repeating sources could simply have undetected subsequent bursts. To address this, we develop a semi-supervised learning framework to identify distinguishing features of repeaters using primary observational parameters from the Blinkverse database, which draws from the CHIME/FRB Catalogs. The framework combines labeled data (known repeaters and confidently classified non-repeaters) with unlabeled sources previously flagged as non-repeaters but exhibiting repeater-like characteristics. We employ uniform manifold approximation and projection with a nearest-neighbor scheme to select potential candidates, followed by semi-supervised classification using five base estimators, including random forest, support vector machine, logistic regression, AdaBoost, and Gradient boost. Each model is fine-tuned through cross-validation, and a voting strategy among the five models is employed to enhance robustness. All models achieve consistently high performance, identifying dispersion measure, peak frequency, and fluence as the most discriminative features. Repeaters tend to show lower dispersion measures, higher peak frequencies, and higher fluences than non-repeaters. We also identify a set of candidate repeaters, several of which are consistent with prior independent studies. Our approach can identify 36 additional repeater candidates that conventional methods may have missed. Finally, the results highlight dispersion measure as a key discriminator between repeaters and non-repeaters, revealing a tension between physical and instrumental origins--either environmental effects, if the two populations arise from distinct progenitors, or detection bias, as nearby sources are more easily observed.

\end{abstract}

\begin{keywords}
methods: data analysis –fast radio bursts.
\end{keywords}



\section{Introduction}


Fast Radio Bursts (FRBs) are extremely bright, millisecond-duration pulses of radio emission first discovered in 2007 \citep{Lorimer2007}. The burst, known as the "Lorimer Burst", exhibited a high dispersion measure (DM) inconsistent with the expected contribution from the Milky Way, suggesting an extragalactic origin. Subsequent discoveries, particularly by \citet{Thornton2013}, confirmed the existence of a population of such bursts, each characterized by large DMs, short durations (typically a few milliseconds), and high fluence (the burst’s total received energy), implying a compact and energetic origin at cosmological distances. The development of wide-field radio telescopes and real-time detection pipelines, such as those employed by CHIME/FRB, ASKAP, and FAST, has led to a dramatic increase in FRB detections in recent years \citep{CHIME2019, Shannon2018, Li2021}, revealing a diverse range of spectral and temporal properties among FRBs. 

One of the most significant breakthroughs in FRB research came with the discovery of the first repeating source, FRB 20121102A, by \citet{Spitler2016}. The source has exhibited hundreds of bursts, often displaying complex frequency structures, downward-drifting subbursts, and temporal clustering, which challenges earlier assumptions that all FRBs were cataclysmic one-off events. Identification of additional repeaters, such as FRB 20180916B and others from CHIME/FRB, has strengthened the notion that repetition is not unique \citep{CHIME2020}. However, many FRBs have not been observed to emit a second burst during follow-up campaigns, leading to their classification as apparently non-repeating or one-off events \citep{Caleb2016, Petroff2016, Bhandari2018}. The dichotomy between repeating and non-repeating FRBs has led to the hypothesis that there may be multiple progenitor types -- such as magnetars, stellar-mass compact object mergers, or stellar-mass compact objects undergoing super-Eddington accretion (the latter known as ultraluminous X-ray sources; ULXs), as well as neutron star collapse -- each producing different emission behaviors \citep{ Petroff2019, zhang2020, sridhar2021}. Alternatively, the observed diversity could result from observational biases, such as beaming effects \citep{connor2020}, burst energy distributions, or limitations due to follow-up sensitivity \citep{kirsten2024} and follow-up duration \citep{ yamasaki2024}. Understanding the nature of FRB repetition is thus central to unraveling their astrophysical origins, constraining source environments, and utilizing FRBs as cosmological probes.

Understanding the relationship between repeaters and non-repeaters requires sophisticated approaches that account for classification uncertainty. Recent studies have explored machine learning (ML) techniques to differentiate repeaters from non-repeaters based on their observed properties. \cite{xu2023} applies a random forest classifier to FRB events in Blinkverse, a database of observations from telescopes such as FAST, CHIME, GBT, and Arecibo. Among five input features used for training, they found frequency bandwidth and fluence to be the most important for classification. \citet{luo2023} applied 6 supervised learning methods to the first CHIME/FRB catalog, using more additional features, and achieved an efficiency of 71–82\% in identifying 27 hidden repeaters.

However, the limitation of supervised learning is label contamination in the training data, as apparent non-repeaters may actually be hidden repeaters, which may lead to mislabeling in the training. In other words, sources classified as non-repeaters after only a single detected burst may actually be repeaters, with additional bursts yet to be observed. \citet{james2023} modeled the DM distribution of FRBs as a function of redshift and found that at least half of the bursts in CHIME’s first catalog may originate from intrinsic repeaters. Likewise, \citet{yamasaki2024} showed that, after accounting for observational biases, the apparent non-repeater source count rate continues to decline over time, revealing a hidden repeater population. Their source-count modeling indicates an intrinsic repeater fraction of at least 50\%, suggesting that many repeaters remain undetected, e.g. because their additional bursts lie below the telescope’s sensitivity or has long waiting times between bursts. Therefore, supervised learning methods often oversimplify this complex problem through strict binary labels. 

Recent studies have used unsupervised ML approaches to identify distinct clusters within the FRB population based on observed properties. These clusters suggest potential subclasses that may be linked to repeating behavior. \citet{chen2022} utilized Uniform Manifold Approximation and Projection (UMAP) to identify 188 candidate repeater sources in the first CHIME/FRB catalog. \citet{zhu2023} applied K-means and Hierarchical Density-Based Spatial Clustering of Applications with Noise (HDBSCAN) algorithms on the same catalog, using nine features, and identified 29 potential hidden repeaters. \citet{yang2023} applied the UMAP and t-distributed stochastic neighbor embedding (t-SNE) to spectrogram representations, revealing a clear visual separation between repeater and non-repeater clusters. \citet{Sharma2024} adopted a semi-supervised positive–unlabeled learning framework with 13 features, identifying 66 repeater candidates from the first CHIME/FRB catalog and the CHIME/FRB Collaboration 2023 catalog. More recently, \citet{Qiang2025} introduced a hybrid unsupervised approach, which identified 270 hidden repeater candidates using the same two catalogs.

Here, we analyze data from the CHIME/FRB Catalog 1 using a semi-supervised learning approach that integrates labeled and unlabeled data. This framework enables us to utilize all available information while considering the possibility that some apparent non-repeaters may actually be unidentified repeaters, thus treating them as unlabelled first. Our method combines UMAP for dimensionality reduction with a nearest neighbors approach to identify unlabeled sources. For the first time, we incorporate a self-training algorithm that iteratively updates the training set by adding sources with high-confidence predictions. Specifically, sources with predicted probabilities exceeding a defined threshold are labeled and merged with the labeled set for retraining in each iteration. The implemented self-training algorithm offers greater adaptability and efficiency compared to previous approaches, making it a robust tool for uncovering hidden repeating FRB sources.

The rest of this paper is organized as follows. Section 2 provides a detailed description of the dataset, including selection criteria, preprocessing steps, and features used for classification. In Section 3, we introduce our semi-supervised framework, highlighting the self-training algorithm and model architecture used to distinguish repeating FRBs from apparently non-repeating ones. Section 4 reports the model performance. We also compare the predictions of our model to those reported in previous literature. The discussion and conclusion are presented in Section 5.

\section{Data and Features}

We use the Blinkverse\footnote{\url{https://blinkverse.zero2x.org/}} database (accessed on 05 April 2025), restricting our selection to data from the CHIME telescope in order to avoid biases arising from heterogeneous instrumental limitations. The resulting subset contains 593 bursts, comprising 137 bursts from known repeaters (originating from 42 repeater sources, with each burst treated as an independent event) and 456 bursts from sources classified as non-repeaters, which are primarily drawn from Catalog 1 and Catalog 2023 of \citet{Amiri2021} and \citet{chime2023}.


From the Blinkverse database \citep{xu2023}, we extract five primary features for each burst to serve as inputs for the training ML model. We follow the primary feature selection of \cite{luo2023}, as these parameters are uniformly available across the dataset. Analyses involving additional features could be deferred to future work. The selected features are dispersion measure while maximizing signal-to-noise ratio ($D_{snr}$), peak flux density in Jy ($F_{d}$), pulse width in ms ($w_p$), peak frequency in MHz ($f_p$), burst fluence in Jy·ms ($f_{lu}$). Note that $D_{snr}$ is the dispersion measure, which measures the total density of free electrons along the line of sight, calculated while maximizing the signal-to-noise ratio of the burst. $F_d$ represents the burst-peak flux as determined from calibrated intensity measurements at maximum signal. $w_p$ indicates the burst duration based on the full width at half maximum (FWHM) of its de-dispersed profile, while $f_p$ marks the frequency at which the burst spectral intensity is maximum. Finally, $f_{lu}$ represents the total received energy per unit area, calculated by integrating the flux density over the burst duration. Due to the observational limitations of the CHIME telescope, which operates within the 400–800~MHz frequency range, the $f_{p}$ values of individual sources fall within the boundaries of this range. However, only 38 cases ($\sim$6\%) have $f_{p}$ values outside the telescope’s coverage, i.e., below 400~MHz or above 800~MHz. For these sources, the $f_{p}$ values are set to the corresponding boundary (400 MHz or 800 MHz).

For an initial statistical assessment, we apply the Mann-Whitney U test to evaluate whether each feature differs significantly between the two groups. This non-parametric test compares independent samples when the dependent variable is ordinal or continuous but not normally distributed. Features with p-value less than 0.01 are considered statistically significant discriminators between repeaters and non-repeaters. The test results are summarized in Table~\ref{tab:repeater_stats}. It can be seen that, nearly all features including $D_{snr}$, $F_{d}$, $w_{p}$, and $f_{p}$ demonstrate statistically significant differences between repeaters and non-repeaters. However, $f_{lu}$ shows no significant differences. The distribution of $D_{snr}$ for repeaters and non-repeaters has previously been reported by \cite{chime2023}, showing results consistent with ours. Note that we use $D_{snr}$ rather than the extragalactic dispersion measure ($eDM$). However, \cite{chime2023} found that both $D_{snr}$ and $eDM$ show statistically significant differences between repeaters and apparently non-repeating sources, so the specific dispersion measure choice ($D_{snr}$ or $eDM$) is not expected to strongly affect the classification results. Note that the results in Table~\ref{tab:repeater_stats} are based on a standard statistical test of the original dataset and do not account for the possibility that some non-repeaters may be hidden repeaters; therefore, the p-value may still not reflect the true level of significant discrimination.

\begin{table}
    \centering
    \caption{The Mann-Whitney U test of features between non-repeaters and repeaters. $\mu_0$ and $\mu_1$ denote the mean value of non-repeaters and repeaters, respectively.}
    \label{tab:repeater_stats}
    \begin{threeparttable}
    \begin{tabular}{lllc}
    \hline
   Feature & $\mu_0$ & $\mu_1$ & p-value\\
    \hline
    $D_{snr} (\rm pc\cdot cm^{-3})$      & 684.75     & 464.83     & $4.10 \times 10^{-9}$   \\
    $F_{d}$ (Jy) & 1.84       & 1.53       & $6.22 \times 10^{-3}$   \\
    $w_{p}$ (ms)         & 7.76       & 3.41       & $1.60 \times 10^{-7}$   \\
    $f_{p}$ (MHz)   & 505.76     & 513.50     & $9.37 \times 10^{-3}$   \\
    $f_{lu} (\rm Jy \cdot ms)$     & 7.02       & 7.77       & $9.32 \times 10^{-1}$   \\
    \hline
    \end{tabular}
    \begin{tablenotes}
    \end{tablenotes}        
    \end{threeparttable}
\end{table}

\begin{figure}
    \centering
    \includegraphics[width=20pc]{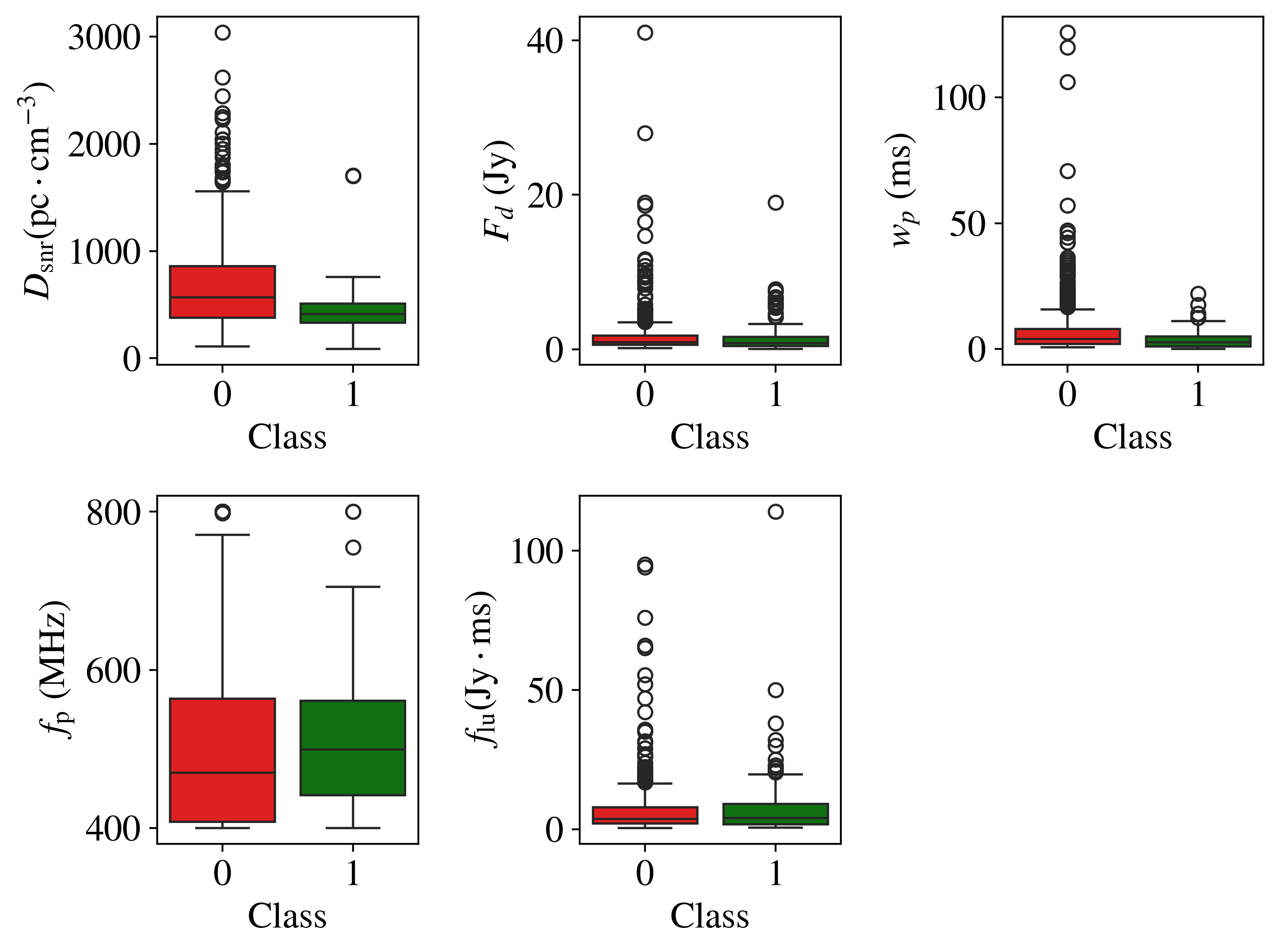}\\
    \caption{Box plots display the distribution of each feature for non-repeating FRBs (class 0) and repeating FRBs (class 1). The box shows the interquartile range (IQR), spanning from the first quartile (Q1) to third quartile (Q3), with a central line marking the median. Whiskers extend to values within 1.5 times the IQR from the quartiles. Points beyond the whiskers are identified as outliers. Note that these data are plotted directly from the original dataset and still do not account for the possibility that some non-repeaters may be hidden repeaters.}
    \label{fig:boxplot}
\end{figure}

Next, we examine the distribution of each feature using box plots, as shown in Fig.~\ref{fig:boxplot}. This visualization reveals the differences in central tendency, spread, and the presence of outliers between the repeating and non-repeating FRB populations. Extreme outliers (i.e., points beyond the whiskers in Fig.~\ref{fig:boxplot}) approximately account for $\sim 5$--12\% of the non-repeater data, and  $\sim 2$--11\% of the repeater data. However, in semi-supervised learning, removing outliers is not always beneficial, since the algorithm uses both labeled and unlabeled data to understand the underlying structure of the dataset. Premature removal of outliers may discard rare but genuine events. Semi-supervised methods can effectively handle such unusual or extreme data points by assigning low-confidence predictions or gradually incorporating them as the model learns and the data structure becomes clearer. Therefore, we retain these outliers in our semi-supervised self-training approach, allowing the model to represent the full data range without bias toward common patterns.


\section{Semi-Supervised Models}

\begin{figure*}
    \centering
    \includegraphics[width=44pc]{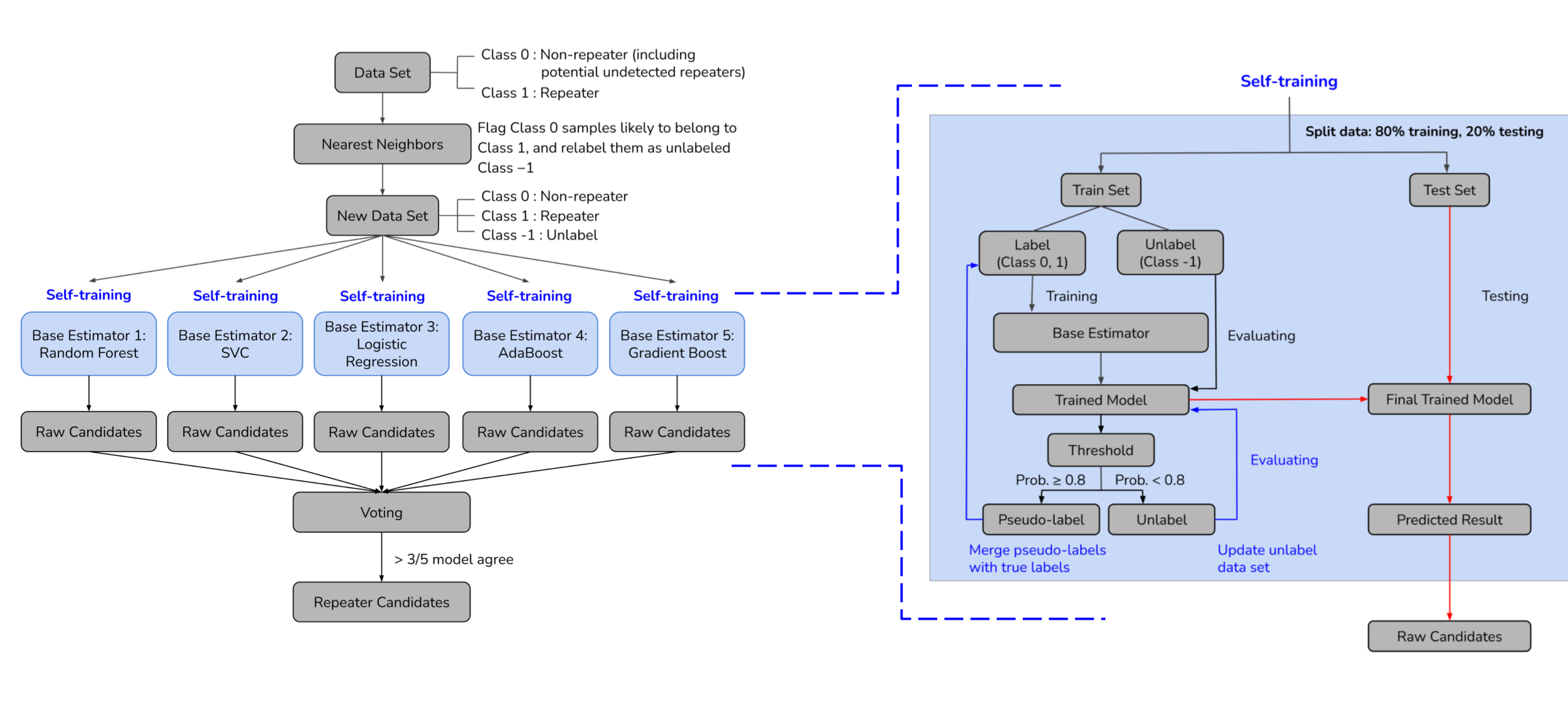}\\
    \caption{Workflow of the semi-supervised learning pipeline using self-training. The initial dataset comprises two classes: Class 0 (non-repeaters) and Class 1 (repeaters). To identify ambiguous non-repeating sources that exhibit repeater-like characteristics, a UMAP-based nearest neighbors approach is applied. These selected Class 0 instances are reassigned to an unlabeled category (Class -1), resulting in a dataset with three classes: 0, 1, and -1. The data is then split into 80\% for training and 20\% for testing. The labeled portion of the training set (Classes 0 and 1) is used to train five separate base classifiers with five features, each within its own self-training algorithm (blue boxes). Each trained model predicts class probabilities for the unlabeled data. Instances with a predicted probability $\geq 0.8$ are assigned pseudo-labels and merged with the labeled set for retraining, while those below the threshold remain unlabeled (blue arrows). This iterative self-training process continues until either all unlabeled samples are assigned or a maximum iteration limit is reached. The final model from each self-training loop is evaluated on the held-out test set to measure performance and to detect raw repeater candidates (red arrows). A voting process is then applied, wherein a source is recognized as a repeater candidate if at least three out of five base estimators predict it as such. See text for more details.}
    \label{fig:workflow}
\end{figure*}

Semi-supervised learning is a machine learning technique that combines elements of supervised learning, which relies on labeled data, and unsupervised learning, which utilizes unlabeled data. In this context, the labeled data consists of known repeaters-FRBs that have emitted two or more bursts and are thus confirmed as repeating sources. The unlabeled data comprises the so-called non-repeaters, which have only been observed to burst once but may emit additional bursts in the future. Therefore, the non-repeater class is inherently uncertain, as it may contain both true non-repeaters and repeaters that have yet to be observed bursting again. Fig.~\ref{fig:workflow} illustrates the overall semi-supervised learning pipeline with self-training used in this study, which is explained step by step below.


\subsection{Data Relabeling}

The primary dataset is initially divided into two classes: non-repeaters (designated as Class 0) and repeaters (Class 1). However, Class 0 may contain a mixture of true non-repeaters and yet-unobserved repeaters, introducing uncertainty into the classification. To address this ambiguity, we implement a relabeling procedure using the nearest neighbors algorithm within a feature space constructed by UMAP, which is a non-linear dimensionality reduction technique that preserves both local and global structures in the data. It excels at clustering and neighborhood-based methods by maintaining the relative distances between nearby data points, which reveals underlying patterns and improves class separation. We use the \texttt{umap-learn} library \citep{mcinnes2018} to perform dimensionality reduction, embedding the data from five features into a two-dimensional space using \texttt{n\_components=2}, \texttt{n\_neighbors=15}, and \texttt{min\_dist=0.1}, with all other parameters set to their default values. The embedding quality and stability for the choice of these hyperparameters are discussed in Appendix~\ref{sec:appendixA}. \texttt{n\_neighbors} controls the size of the local neighborhood considered when constructing the low-dimensional map. Lower values emphasize the preservation of fine-grained local structures, while higher values favor the retention of broader global relationships. The parameter \texttt{min\_dist} regulates how tightly points are allowed to cluster in the low-dimensional embedding, with smaller values producing more compact clusters and larger values yielding more diffuse distributions. By reducing the data to two dimensions, UMAP allows for better neighbor identification and clearer visual analysis.

Within the UMAP-embedded space, we apply the nearest neighbors algorithm \citep{cover1967} to each instance originally labeled as Class 0. This is done using the \texttt{sklearn.neighbors.NearestNeighbors} \citep{scikit-learn} algorithm with the hyperparameter \texttt{n\_neighbors=11}, keeping all other parameters at their default values. Distances are computed to identify the nearest neighbors for each data point of Class 0, yielding 10 surrounding neighbors per instance (excluding the point itself). For each instance, we compute the proportion of its nearest neighbors that are labeled as repeaters (Class 1). If this proportion exceeds a predefined threshold, empirically set to 20\% in our study, the instance is marked as a potential repeater candidate. These instances are then relabeled as unlabeled data (Class -1), allowing the semi-supervised model to treat them as uncertain cases first, rather than directly assigning them to the non-repeater class. After this relabeling process, the dataset is divided into three categories: Class 0 (non-repeaters), Class 1 (repeaters), and Class -1 (unlabeled).

\subsection{Based estimators and Hyperparameter tuning}

We randomly split the full dataset, allocating 80\% for training and 20\% for testing. The training set contains both labeled data (Class 0 and Class 1) and unlabeled data (Class -1) from our earlier relabeling process. The dataset contains 217 instances of Class 0, 137 of Class 1, and 239 of Class -1, resulting in an imbalance between Class 0 and Class 1. To mitigate biased predictions and reduce the risk of overfitting, we apply the synthetic minority over-sampling technique (SMOTE) \citep{chawla2002} using the \texttt{imbalanced-learn} library \citep{lemaavztre2017} to balance the dataset. For each minority class, SMOTE selects neighbors of the original instance, usually using k-nearest neighbors, then creates new synthetic samples along the line between the original sample and its neighbor to increase the number of minority class samples.

We then train the based estimators using only the labeled portion (Class 0 and Class 1) with five features. These based estimators can be any supervised machine learning model that produces probability outputs. Here, we employ five famous models, including random forest (RF), support vector classification (SVC), logistic regression (LR), AdaBoost (Ada), and Gradient boost (GB) to develop initial models trained exclusively on known repeaters (Class 1) and identified non-repeaters (Class 0). The implementations are based on the \texttt{scikit-learn} library \citep{scikit-learn}, corresponding to \texttt{RandomForestClassifier}, \texttt{SVC} with probability output enabled, \texttt{LogisticRegression}, \texttt{AdaBoostClassifier}, and \texttt{GradientBoostingClassifier}, respectively. The outputs from these based models are then aggregated through a voting mechanism which will be explained later.

\begin{table*}
\centering
\caption{Search space for hyperparameters of each base estimator model employed in \texttt{BayesSearchCV}.}
\label{tab:search_space}
\begin{tabular}{|l|l|l|}
\hline
\textbf{Model} & \textbf{Hyperparameters} & \textbf{Search Space} \\ \hline

Random Forest Classifier 
& \texttt{n\_estimators} & 50 to 500 (integer) \\
& \texttt{min\_samples\_split} & 2 to 40 (integer) \\
& \texttt{min\_samples\_leaf} & 1 to 40 (integer) \\ \hline

Support Vector Classifier 
& \texttt{C} & $10^{-3}$ to $10^{3}$ (log-uniform) \\
& \texttt{degree} & 1 to 10 (integer) \\ \hline

Logistic Regression 
& \texttt{tol} & $10^{-5}$ to $10^{-3}$ (log-uniform) \\
& \texttt{C} & $10^{-3}$ to $10^{3}$ (log-uniform) \\
& \texttt{solver} & lbfgs, liblinear, newton-cg, newton-cholesky, sag, saga\\ \hline

AdaBoost Classifier 
& \texttt{n\_estimators} & 50 to 500 (integer) \\
& \texttt{learning\_rate} & 0.001 to 2.0 (log-uniform) \\ \hline

Gradient Boosting Classifier 
& \texttt{n\_estimators} & 50 to 500 (integer) \\
& \texttt{learning\_rate} & 0.01 to 1.0 (log-uniform) \\
& \texttt{max\_depth} & 3 to 10 (integer) \\ \hline

\end{tabular}
\end{table*}

We employ \texttt{BayesSearchCV} from the \texttt{scikit-optimize} library \citep{head2021} to fine-tune the model hyperparameters. Unlike traditional grid or random search methods, \texttt{BayesSearchCV} strategically selects parameter combinations that are most likely to improve model performance by leveraging information from previous evaluations. This approach balances exploration of high-uncertainty regions in the parameter space with exploitation of areas that have shown promising results. The specific hyperparameters tuned for each ML model are summarized in Table~\ref{tab:search_space}. 

For the RF, we tune the number of trees (\texttt{n\_estimators}), the minimum number of samples required to split an internal node (\texttt{min\_samples\_split}), and the minimum number of samples required at a leaf node (\texttt{min\_samples\_leaf}). For SVC, we tune the regularization strength (\texttt{C}) that balances between maximizing the margin and minimizing classification errors. The polynomial degree (\texttt{degree}) is used when the polynomial kernel is applied. For LR, we tune the regularization strength \texttt{C}, the tolerance (\texttt{tol}) that sets the stopping criterion for optimization, and the \texttt{solver} that defines the optimization algorithm. For Ada, the number of boosting stages (\texttt{n\_estimators}) determines how many weak learners are combined, and the learning rate (\texttt{learning\_rate}) scales the contribution of each learner. Finally, in the GB, the number of boosting iterations (\texttt{n\_estimators}), learning rate (\texttt{learning\_rate}), and the maximum depth of individual trees (\texttt{max\_depth}) are tuned. These parameters control model complexity and help prevent overfitting. The rest of the parameters in each base estimator are set as default.

\subsection{Self-training process}

Once each base estimator is trained, it is applied to the unlabeled samples (Class $-1$) to estimate their class probabilities, as illustrated by the blue line in Fig.~\ref{fig:workflow}. For each unlabeled instance, the model predicts a class label—either Class 0 or Class 1—along with an associated probability. If the predicted class probability exceeds a confidence threshold of 80\%, the instance is assigned a pseudo-label corresponding to the predicted class. Otherwise, it remains unlabeled.

The newly pseudo-labeled samples are combined with the original labeled data to create an expanded training set, which is used to retrain the base estimator. The updated model then predicts classes for the remaining unlabeled instances. This iterative process, shown by the blue line in Fig.~\ref{fig:workflow}, forms a self-training loop that continues until either all unlabeled data points receive pseudo-labels or the model reaches a maximum number of iterations. The maximum number of iterations was set to 100, and we have verified that the solution has converged.

\subsection{Testing process, Accuracy measurements and voting}

Multiple iterations of the self-training process and pseudo-labeling approach in previous steps help refine the model-decision boundaries and increase its confidence in classifying ambiguous cases where the distinctions may be subtle. Finally, the fully trained model undergoes evaluation on the held-out test set, depicted by the red lines in Fig.~\ref{fig:workflow}. Based on this workflow, the predictions are derived exclusively from instances within the test set. To ensure that each source appears in the test set, the data is randomly split 1,000 times, and the entire process after that is repeated accordingly. For sources initially labeled as Class 0 and Class -1 in the test set, those predicted as Class 1 in over 50\% of their test-set appearances are considered Class 1 raw candidates. To increase robustness, we require that at least three of the five estimators agree on the classification of a raw candidate. If this condition is met, the source is considered a confirmed candidate.

In binary classification, model performance is evaluated using several key indicators. True Positives (TP) count instances correctly identified as positive, such as accurately detecting a repeater. False Positives (FP) represent instances incorrectly labeled as positive (e.g. classifying a non-repeater as a repeater). False Negatives (FN) occur when the model mistakenly identifies positive instances as negative, failing to detect true repeaters. To assess its classification performance, standard evaluation metrics including precision, recall, and $F_{1}$ score are calculated, with equations given, respectively, below 
\begin{equation}
\text{Precision} = \frac{\text{TP}}{\text{TP} + \text{FP}} \;,
 \label{eq:Precision}
\end{equation}
\begin{equation}
\text{Recall} = \frac{\text{TP}}{\text{TP} + \text{FN}} \;,
 \label{eq:Recall}
\end{equation}
\begin{equation}
F_{1} = 2 \cdot \frac{\text{Precision} \cdot \text{Recall}}{\text{Precision} + \text{Recall}}\;.
 \label{eq:F1}
\end{equation}
While the precision measures the proportion of predicted positive instances that are actually positive, the recall measures the proportion of actual positive instances that are correctly identified by the model. The $F_{1}$ score is the harmonic mean of the precision and recall, providing a balanced measure when both metrics are equally important.

\subsection{Feature important}

We also utilize SHAP (SHapley Additive exPlanations) \citep{lundberg2017} to assess feature importance for each ML model. SHAP leverages principles from cooperative game theory to determine how much each feature contributes to a machine learning output. SHAP values are computed using \texttt{TreeExplainer} \citep{lundberg2020} and \texttt{KernelExplainer} functions available in the SHAP  Python library, based on the average marginal contribution of a feature across all subsets of features. This approach ensures a fair and consistent measure of feature importance.

We also examine feature effects using Partial Dependence Plots (PDPs) \citep{hastie2009}, computed with the \texttt{sklearn.inspectionpartial\_dependence} algorithm \citep{scikit-learn}. This function estimates the average predicted response of the model as one or two features vary, while averaging over all other features. PDPs illustrate the marginal effect of selected features on the prediction, helping to reveal whether their relationship with the target is linear, monotonic, or more complex \citep{friedman2001}.


\section{Results}

\begin{figure}
    \centering
    \begin{subfigure}{0.5\textwidth}
        \includegraphics[width=\linewidth]{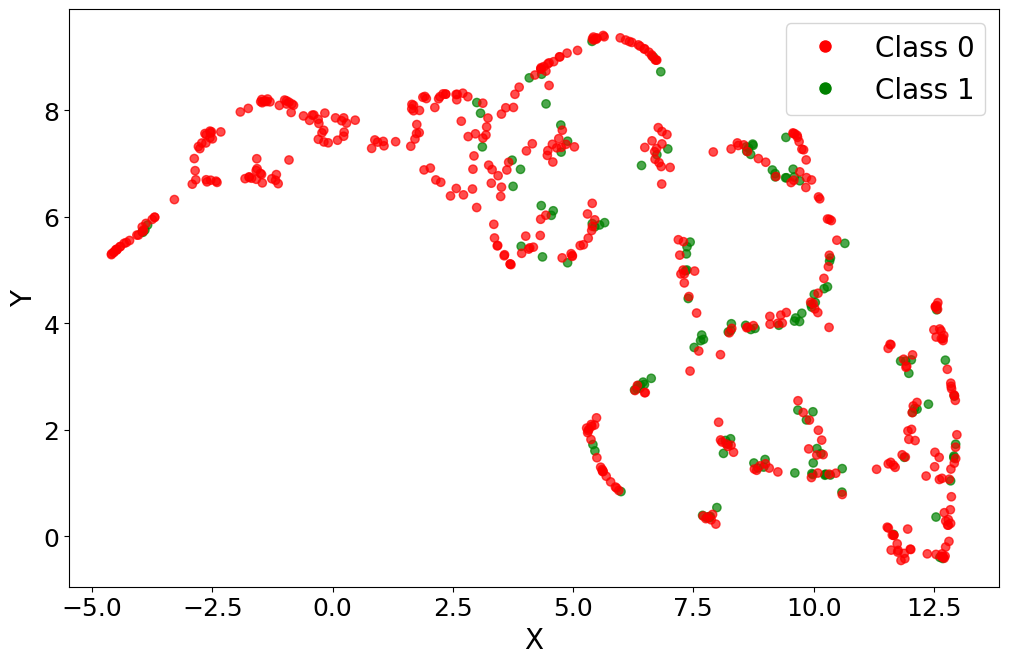}
    \end{subfigure}%
    \hfill
    \begin{subfigure}{0.5\textwidth}
        \includegraphics[width=\linewidth]{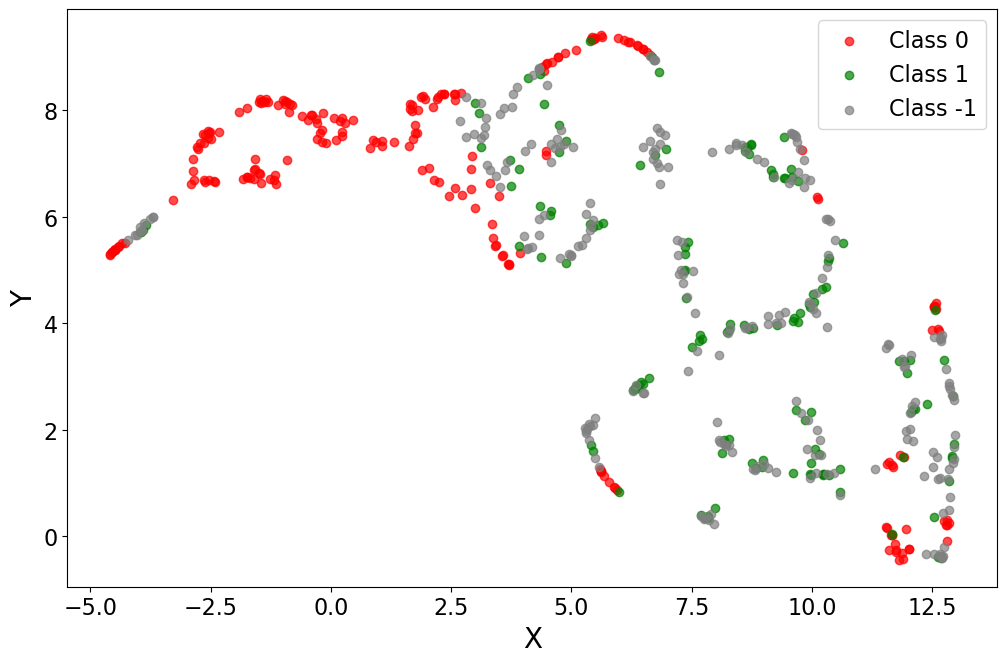}
    \end{subfigure}
    \caption{UMAP visualization of the dataset. The top panel displays the original label distribution, with repeaters (Class 1) shown in green and non-repeaters (Class 0) in red. The bottom panel shows the updated labels after neighborhood-based relabeling, where some Class 0 instances exhibiting repeater-like characteristics are reassigned to the unlabeled category (Class -1), shown in gray.}
    \label{fig:Umap_1}
\end{figure}

At the first step, we investigate all data points in the UMAP feature space using a nearest-neighbor approach. The results before and after relabeling are shown in Fig.~\ref{fig:Umap_1}. Note that non-repeaters (Class 0) that have at least 20\% (i.e., 2 out of 10) of their 10 nearest neighbors belonging to known-repeater (Class 1) are reassigned to the unlabeled class (Class -1). Using this criterion, 239 out of 456 non-repeaters are relabeled as unlabeled. Increasing this threshold to 30\% reduces the number of relabeled samples, but it also limits the opportunity to capture ambiguous cases that could benefit from self-training. Conversely, lowering the threshold to 10\% increases the number of unlabeled samples, yet the low probability of true ambiguity may lead to higher misclassification rates. Nevertheless, We additionally tested thresholds of 10\% and 30\% to assess downstream performance. Among these, the 20\% threshold offers the most effective compromise and achieves the highest performance (see Appendix~\ref{sec:appendixB}). We therefore adopt the 20\% threshold as a balance between capturing uncertain cases and minimizing classification errors.

The average efficiency of the five base estimators on the test set is high, with each achieving a recall score, precision, and $F_{1}$-score of at least 0.8, as shown in Table~\ref{tab:model_performance}, indicating strong classification performance. This demonstrates the robustness of the voting process. Among the models, the Ada model achieved the best performance, with all recall, precision, and $F_{1}$ scores reaching 0.9. The corresponding training scores vary by only 3--5\% from the test scores, suggesting that the models are well-generalized and not overfitting.

\begin{table}
\centering
\caption{Comparison of average testing performance scores across different base estimator models.}
\begin{tabular}{lccc}
\hline
Model & Recall score & Precision & $F_1$ score \\
\hline
Random Forest          & 0.85 & 0.86 & 0.85 \\
Support Vector Machine & 0.83 & 0.82 & 0.83 \\
AdaBoost               & 0.90 & 0.90 & 0.90 \\
Logistic Regression    & 0.82 & 0.90 & 0.86 \\
Gradient Boosting      & 0.84 & 0.88 & 0.86 \\
\hline
\end{tabular}
\label{tab:model_performance}
\end{table}

\begin{figure}
    \centering
    \includegraphics[width=20pc]{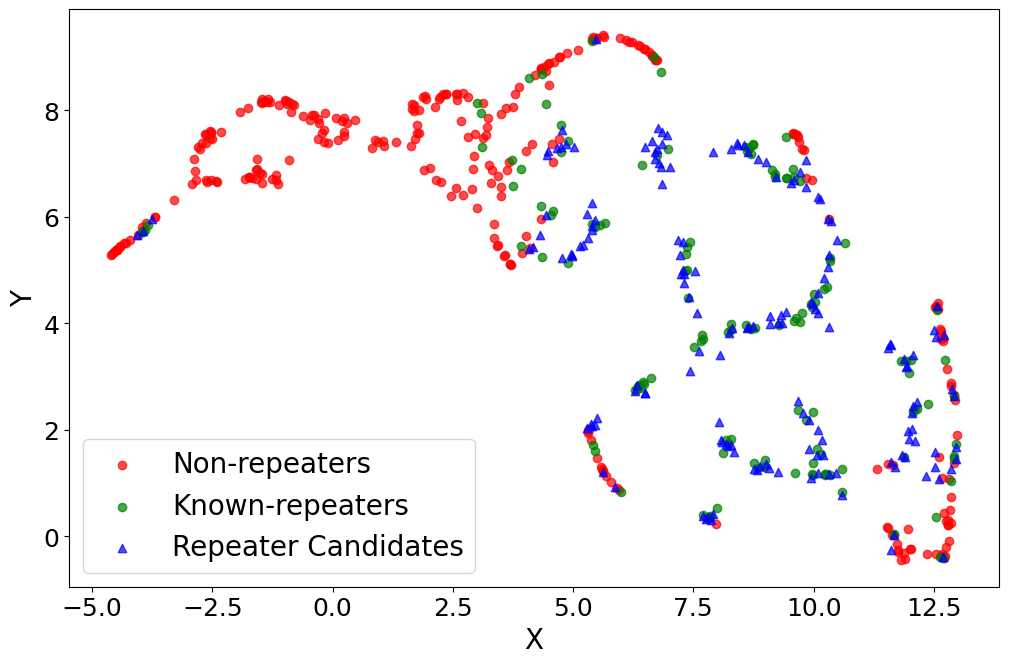}\\
    \caption{UMAP visualization highlighting our repeater candidates. Red circle, green circle, and blue triangle represent non-repeaters, known repeaters, and repeater candidates identified in this study, respectively.}
    \label{fig:Umap_candidate}
\end{figure}

Our candidate selection was based on a class probability threshold of $\geq 0.8$ during self-training, combined with a voting scheme that required agreement from at least three out of five models to ensure robustness, as described in Section 3.3 and 3.4. Using this approach, the model identified 168 high-confidence repeater candidates from the previously non-repeater (Class 0) and unlabeled (Class -1) sets. Of these, 153 candidates originated from the unlabeled class, and the remaining 15 came from the non-repeater class, meaning that $\sim 91\%$ of all new repeater candidates were drawn from the unlabeled class. This strong proportion suggests that the unlabeled set is an effective source for discovering potential repeaters. The distribution of the new repeater candidates in UMAP space is shown in Fig.~\ref{fig:Umap_candidate}. As expected, they cluster around known repeaters in the feature space.

For comparison, we also examine a baseline model in which the unlabeled class (-1) is not introduced. In this case, only classes 0 and 1 are used  for self-training to obtain the raw candidates, without any iteration, while all other aspects of the method remain unchanged. This provides a direct baseline for assessing the impact of introducing class -1 under a consistent experimental setup.

Table~\ref{tab:longcandidate} lists our predicted repeaters, along with their CHIME/FRB catalog IDs, voting results, cross-matches with previous studies, and comparison against the baseline model. Among the 168 candidates, 140 were supported by at least four base estimators, and 89 were unanimously supported by all five estimators.Cross-matching with five previous studies \citep{chen2022, luo2023, zhu2023, Sharma2024, Qiang2025} reveals that 11 sources are consistent with all five studies, 16 match four studies, 16 match three studies, 32 match two studies, and 61 match one study. In total, 136 candidates correspond to at least one previous study, while 32 are newly identified repeater candidates that were not reported previously. We identified 13 high-confidence candidates that were supported by at least four base estimators and matched with at least four previous studies. Finally, our results suggest the highest-confidence group consists of sources supported by all five base estimators and matched across all previous studies. These sources are FRB20190429B, FRB20181221A, FRB20181017B, FRB20190609A, FRB20190112A.

Nevertheless, the baseline model identifies 132 candidates, whereas introducing the ambiguity class (-1) increases this to 168, yielding 36 additional candidates ($\sim 27$\%). This highlights the clear benefit of incorporating the ambiguity class into the self-training procedure. We additionally tested neighborhood-fraction thresholds of 10\% and 30\%. Under these settings, the baseline model identifies 218 and 94 candidates, respectively, whereas the semi-supervised approach recovers 68 and 23 additional repeater candidates beyond these baselines. Importantly, across all thresholds examined, the baseline model never discovers candidates that are missed by the semi-supervised model with the ambiguity class (-1), demonstrating that the self-training framework consistently expands, rather than reshuffles, the candidate pool. Despite these variations, the 20\% threshold remains the most suitable choice, as it delivers the strongest downstream performance (see also Appendix~\ref{sec:appendixB}).

\onecolumn
\begin{longtable}{llll}  
\caption{Repeater candidates identified by model voting and cross-matched with previous studies. Note that a source is identified as a repeater candidate when at least three out of the five models predict it as a repeater. C22, L23, Q25, S24, and Z23 represent \citet{chen2022}, \citet{luo2023}, \citet{zhu2023}, \citet{Sharma2024}, and \citet{Qiang2025}, respectively. In comparison with the baseline model,  ‘Y’ indicates that the baseline model also identifies the source as a repeater. ‘N’ indicates that the baseline model does not recover the source, meaning the source is recovered only after introducing class -1 in the self-training process.} 

\label{tab:longcandidate} \\
\toprule
\textbf{Source} & \textbf{Model Voted} & \textbf{Matched Works} & \textbf{Baseline Model} \\
\midrule
\endfirsthead

\toprule
\textbf{Source} & \textbf{Model Voted} & \textbf{Matched Works} & \textbf{Baseline Model} \\
\midrule
\endhead

\midrule
\multicolumn{3}{r}{\textit{Continued on next page}} \\
\midrule
\endfoot

\bottomrule
\endlastfoot

FRB20180727A & Ada, GB, RF, SVC & - & Y \\
FRB20180729B & Ada, GB, LR, RF, SVC & C22, Q25 & Y \\
FRB20180810A & Ada, GB, LR, RF, SVC & - & Y \\
FRB20180810B & Ada, GB, LR, RF & Q25 & Y \\
FRB20180814B & Ada, LR, RF & - & N \\
FRB20180904A & Ada, GB, LR, RF, SVC & Q25 & Y \\
FRB20180907E & Ada, GB, LR, RF & C22, S24, Q25 & Y \\
FRB20180915A & Ada, GB, LR, RF, SVC & Q25 & Y \\
FRB20180916A & Ada, GB, LR, RF, SVC & Q25 & N \\
FRB20180919B & GB, LR, RF, SVC & - & Y \\
FRB20180920A & Ada, GB, RF, SVC & C22, Q25 & Y \\
FRB20180920B & Ada, GB, RF, SVC & C22, S24, Q25 & Y \\
FRB20180921A & Ada, GB, LR, RF, SVC & Q25 & N \\
FRB20180923D & LR, RF, SVC & C22, Q25 & N \\
FRB20181013A & GB, LR, RF, SVC & Q25 & N \\
FRB20181014D & Ada, GB, LR, RF, SVC & C22 & Y \\
FRB20181017B & Ada, GB, LR, RF, SVC & C22, L23, Z23, S24, Q25 & Y \\
FRB20181018B & Ada, GB, LR, RF, SVC & Q25 & Y \\
FRB20181022C & Ada, GB, LR, RF, SVC & Q25 & Y \\
FRB20181022D & Ada, GB, LR, RF, SVC & C22, Q25 & Y \\
FRB20181022E & Ada, GB, LR, RF, SVC & C22, Q25 & N \\
FRB20181030C & GB, LR, RF, SVC & Z23, Q25 & N \\
FRB20181030D & Ada, GB, LR, RF, SVC & Q25 & Y \\
FRB20181104C & GB, LR, RF, SVC & - & Y \\
FRB20181116A & Ada, GB, LR, RF, SVC & Q25 & Y \\
FRB20181117B & Ada, GB, LR, RF, SVC & Q25 & Y \\
FRB20181118A & Ada, GB, LR, RF, SVC & C22 & Y \\
FRB20181118B & Ada, GB, LR, RF, SVC & - & Y \\
FRB20181119B & GB, LR, RF, SVC & Q25 & Y \\
FRB20181119C & GB, LR, RF & Q25 & N \\
FRB20181125A & Ada, GB, LR, RF & C22, Z23, S24, Q25 & Y \\
FRB20181126A & Ada, GB, LR, RF, SVC & C22, Q25 & Y \\
FRB20181128B & Ada, GB, LR, SVC & Q25 & N \\
FRB20181128C & Ada, GB, RF, SVC & C22, L23, S24, Q25 & Y \\
FRB20181129A & Ada, GB, LR, RF, SVC & - & Y \\
FRB20181129B & Ada, GB, LR, RF, SVC & C22, S24, Q25 & Y \\
FRB20181129C & Ada, GB, LR, RF, SVC & - & Y \\
FRB20181202A & GB, RF, SVC & - & N \\
FRB20181202C & Ada, GB, RF, SVC & - & Y \\
FRB20181203B & Ada, GB, LR, RF, SVC & C22, S24, Q25 & Y \\
FRB20181213B & GB, RF, SVC & C22, S24, Q25 & Y \\
FRB20181214A & Ada, GB, LR, RF, SVC & C22, Z23, S24, Q25 & Y \\
FRB20181215A & Ada, LR, SVC & Q25 & N \\
FRB20181215B & Ada, GB, LR, RF, SVC & C22 & Y \\
FRB20181216A & Ada, GB, LR, RF, SVC & C22, Q25 & Y \\
FRB20181218C & Ada, GB, LR, RF, SVC & L23, Q25 & Y \\
FRB20181221A & Ada, GB, LR, RF, SVC & C22, L23, Z23, S24, Q25 & Y \\
FRB20181222B & GB, LR, RF & - & N \\
FRB20181222E & Ada, GB, LR, RF, SVC & Q25 & Y \\
FRB20181223B & Ada, GB, RF, SVC & C22, S24, Q25 & Y \\
FRB20181224A & Ada, GB, LR, RF, SVC & Q25 & Y \\
FRB20181225B & Ada, GB, LR, RF, SVC & Q25 & N \\
FRB20181226B & Ada, GB, LR, RF, SVC & Q25 & Y \\
FRB20181226D & LR, RF, SVC & Q25 & N \\
FRB20181228B & Ada, GB, LR, RF, SVC & C22, Z23, S24, Q25 & Y \\
FRB20181229B & Ada, GB, RF, SVC & C22, L23, S24, Q25 & Y \\
FRB20181231B & GB, LR, RF, SVC & C22, L23, Z23, S24, Q25 & Y \\
FRB20190102B & Ada, GB, LR, RF, SVC & Q25 & N \\
FRB20190103B & Ada, GB, RF, SVC & S24, Q25 & Y \\
FRB20190104B & GB, LR, RF & - & N \\
FRB20190106A & GB, LR, RF & C22, L23, S24, Q25 & N \\
FRB20190106B & GB, LR, RF, SVC & C22, Q25 & Y \\
FRB20190109A & Ada, GB, LR, RF, SVC & C22, S24, Q25 & Y \\
FRB20190110B & GB, RF, SVC & - & Y \\
FRB20190112A & Ada, GB, LR, RF, SVC & C22, L23, Z23, S24, Q25 & Y \\
FRB20190116C & GB, LR, RF, SVC & C22 & Y \\
FRB20190116F & Ada, GB, LR, RF, SVC & C22, Q25 & Y \\
FRB20190118A & LR, RF, SVC & C22, Q25 & N \\
FRB20190121A & Ada, GB, LR, RF, SVC & - & Y \\
FRB20190122B & Ada, GB, LR, RF, SVC & C22 & Y \\
FRB20190124B & Ada, GB, LR, RF, SVC & - & Y \\
FRB20190124C & GB, LR, RF, SVC & Q25 & Y \\
FRB20190124D & GB, LR, RF & Q25 & Y \\
FRB20190124E & Ada, GB, RF, SVC & C22, S24, Q25 & Y \\
FRB20190124F & Ada, LR, RF & Q25 & Y \\
FRB20190125A & Ada, GB, RF, SVC & C22, L23, Z23, S24, Q25 & Y \\
FRB20190125B & LR, RF, SVC & C22, L23, Z23, Q25 & Y \\
FRB20190128B & GB, LR, RF, SVC & Q25 & Y \\
FRB20190128C & Ada, GB, LR, RF, SVC & C22, L23, S24, Q25 & Y \\
FRB20190128D & Ada, GB, LR, RF, SVC & Q25 & Y \\
FRB20190129A & Ada, GB, LR, RF, SVC & C22, L23, S24, Q25 & Y \\
FRB20190131C & Ada, GB, LR, RF, SVC & - & Y \\
FRB20190131D & GB, LR, RF, SVC & C22 & Y \\
FRB20190202A & Ada, GB, LR, RF, SVC & Q25 & Y \\
FRB20190202B & GB, LR, RF, SVC & - & Y \\
FRB20190203A & Ada, GB, LR, RF, SVC & C22, Q25 & Y \\
FRB20190203B & Ada, GB, LR, RF, SVC & Q25 & Y \\
FRB20190204A & Ada, GB, LR, RF, SVC & C22, Q25 & Y \\
FRB20190206A & Ada, GB, LR, RF, SVC & C22, L23, S24, Q25 & N \\
FRB20190206B & Ada, GB, RF, SVC & L23, S24, Q25 & Y \\
FRB20190210D & GB, LR, RF, SVC & C22, Q25 & Y \\
FRB20190210E & GB, LR, RF, SVC & C22, Q25 & Y \\
FRB20190211B & Ada, GB, LR, RF, SVC & Q25 & Y \\
FRB20190214C & Ada, GB, LR, RF, SVC & Q25 & N \\
FRB20190218B & Ada, GB, RF, SVC & C22, L23, Z23, S24, Q25 & Y \\
FRB20190218C & Ada, GB, LR, RF, SVC & Q25 & N \\
FRB20190219B & Ada, GB, SVC & C22 & N \\
FRB20190221D & Ada, GB, LR, RF, SVC & Q25 & Y \\
FRB20190222C & Ada, GB, LR, RF, SVC & C22, Q25 & Y \\
FRB20190223A & Ada, GB, LR, RF, SVC & C22, S24, Q25 & Y \\
FRB20190223B & Ada, GB, LR, RF, SVC & C22 & Y \\
FRB20190224C & GB, LR, RF, SVC & C22 & Y \\
FRB20190224E & Ada, GB, LR, RF, SVC & Q25 & Y \\
FRB20190226A & GB, LR, RF, SVC & - & Y \\
FRB20190227A & Ada, GB, LR, RF, SVC & C22 & Y \\
FRB20190227B & GB, LR, RF & Q25 & Y \\
FRB20190228A & Ada, GB, RF, SVC & C22, Z23, S24, Q25 & N \\
FRB20190304A & GB, LR, RF, SVC & C22, Q25 & N \\
FRB20190304C & Ada, GB, LR, RF, SVC & C22, Q25 & Y \\
FRB20190308C & Ada, GB, RF, SVC & C22, Q25 & Y \\
FRB20190309A & Ada, GB, LR, RF, SVC & C22, Q25 & Y \\
FRB20190316A & Ada, GB, LR, RF, SVC & - & Y \\
FRB20190317B & Ada, GB, LR, RF, SVC & - & Y \\
FRB20190317C & GB, RF, SVC & - & Y \\
FRB20190320C & Ada, GB, LR, RF, SVC & Q25 & Y \\
FRB20190320E & Ada, GB, LR, RF, SVC & C22, Q25 & Y \\
FRB20190323C & Ada, GB, LR, RF, SVC & Q25 & Y \\
FRB20190328B & Ada, GB, LR, RF, SVC & - & Y \\
FRB20190403A & Ada, GB, LR, RF, SVC & Z23 & N \\
FRB20190403B & Ada, GB, LR, RF, SVC & Q25 & N \\
FRB20190403E & Ada, GB, LR, RF, SVC & C22, S24, Q25 & Y \\
FRB20190405A & GB, RF, SVC & Q25 & Y \\
FRB20190409A & Ada, GB, SVC & - & N \\
FRB20190409B & Ada, GB, RF, SVC & L23, Z23, S24, Q25 & Y \\
FRB20190410A & Ada, GB, LR, RF, SVC & C22, L23, S24, Q25 & Y \\
FRB20190410B & GB, LR, RF, SVC & C22, Z23, Q25 & Y \\
FRB20190411C & Ada, GB, LR, RF, SVC & C22, S24, Q25 & Y \\
FRB20190412A & Ada, GB, LR, RF, SVC & C22 & Y \\
FRB20190415A & Ada, GB, RF, SVC & - & Y \\
FRB20190415C & GB, RF, SVC & Q25 & N \\
FRB20190417C & Ada, GB, LR, RF, SVC & C22, Q25 & Y \\
FRB20190419A & Ada, GB, LR, RF, SVC & C22, S24, Q25 & N \\
FRB20190422A & Ada, GB, RF, SVC & C22, L23, Z23, S24, Q25 & Y \\
FRB20190423B & GB, RF, SVC & C22, L23, Z23, S24, Q25 & Y \\
FRB20190423D & Ada, GB, LR, RF, SVC & - & Y \\
FRB20190426A & GB, LR, RF, SVC & C22, Q25 & Y \\
FRB20190427A & Ada, LR, SVC & Q25 & N \\
FRB20190429A & Ada, GB, LR, RF, SVC & C22 & Y \\
FRB20190429B & Ada, GB, LR, RF, SVC & C22, L23, Z23, S24, Q25 & Y \\
FRB20190430A & Ada, GB, RF, SVC & C22, S24, Q25 & Y \\
FRB20190502C & Ada, GB, LR, RF, SVC & Q25 & Y \\
FRB20190515A & Ada, GB, RF, SVC & - & Y \\
FRB20190515D & Ada, GB, LR, RF, SVC & C22, Q25 & Y \\
FRB20190517C & Ada, GB, LR, RF, SVC & C22, Q25 & N \\
FRB20190518C & Ada, GB, LR, RF, SVC & Q25 & Y \\
FRB20190518G & GB, LR, RF, SVC & C22, Q25 & Y \\
FRB20190519G & Ada, GB, LR, RF, SVC & - & Y \\
FRB20190520A & Ada, GB, LR, RF, SVC & C22, Q25 & Y \\
FRB20190527A & GB, RF, SVC & C22, L23, Z23, S24, Q25 & N \\
FRB20190527C & Ada, GB, LR, RF, SVC & - & Y \\
FRB20190530A & Ada, GB, LR, RF, SVC & C22, S24, Q25 & Y \\
FRB20190531C & Ada, GB, LR, RF, SVC & C22, Z23, S24, Q25 & Y \\
FRB20190601C & Ada, GB, LR, RF, SVC & C22, Z23, S24, Q25 & N \\
FRB20190604C & Ada, GB, RF, SVC & Q25 & Y \\
FRB20190606B & GB, LR, RF, SVC & Q25 & Y \\
FRB20190609A & Ada, GB, LR, RF, SVC & C22, L23, Z23, S24, Q25 & Y \\
FRB20190609B & GB, LR, RF, SVC & Q25 & Y \\
FRB20190612A & Ada, GB, RF, SVC & - & Y \\
FRB20190614B & GB, LR, RF, SVC & - & Y \\
FRB20190617B & LR, RF, SVC & C22, Z23, S24, Q25 & N \\
FRB20190619B & GB, LR, RF & Q25 & N \\
FRB20190621C & GB, LR, RF, SVC & C22, Q25 & Y \\
FRB20190624B & GB, LR, SVC & C22, Q25 & N \\
FRB20190625A & Ada, GB, LR, RF & Q25 & Y \\
FRB20190627A & GB, LR, RF, SVC & Q25 & Y \\
FRB20190627B & Ada, GB, LR, RF, SVC & - & Y \\
FRB20190629A & Ada, GB, LR, RF, SVC & C22, Q25 & Y \\
FRB20190630C & Ada, GB, SVC & - & Y \\
 \hline


\end{longtable}

\twocolumn





\begin{figure*}
    \centering
    \begin{subfigure}{0.49\textwidth}
        \centering
        \texttt{RF}\\
        \includegraphics[width=\linewidth]{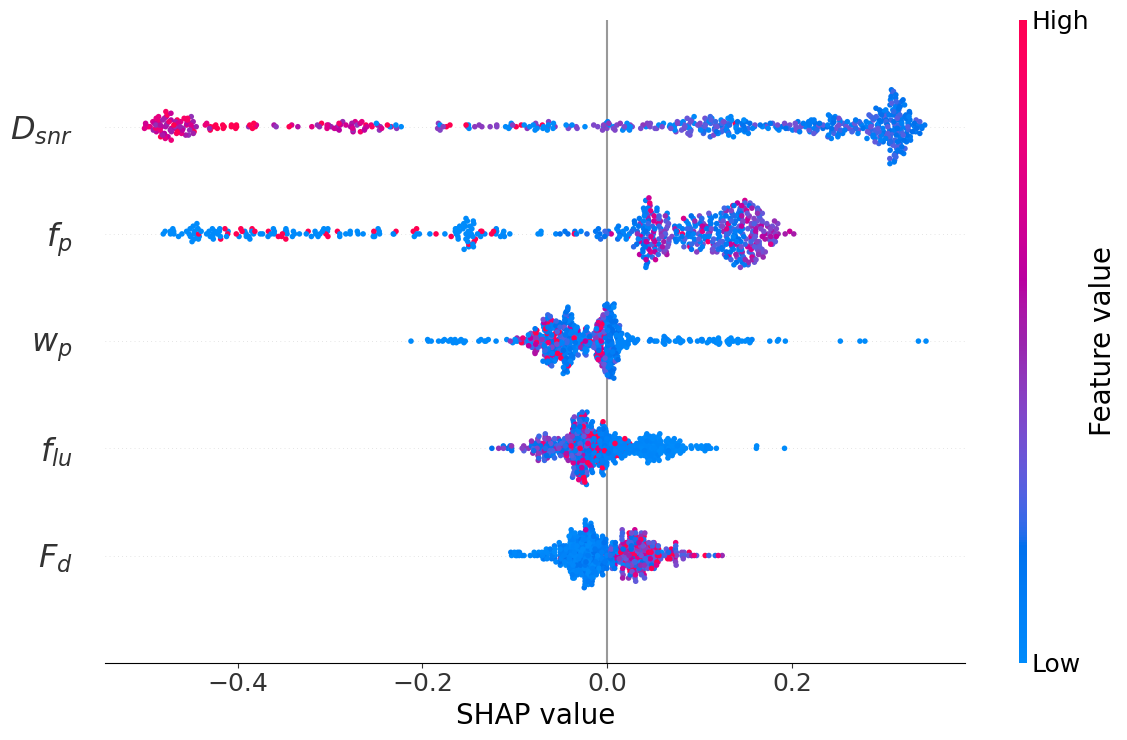}
    \end{subfigure}
    \begin{subfigure}{0.49\textwidth}
        \centering
        \texttt{SVC}\\
        \includegraphics[width=\linewidth]{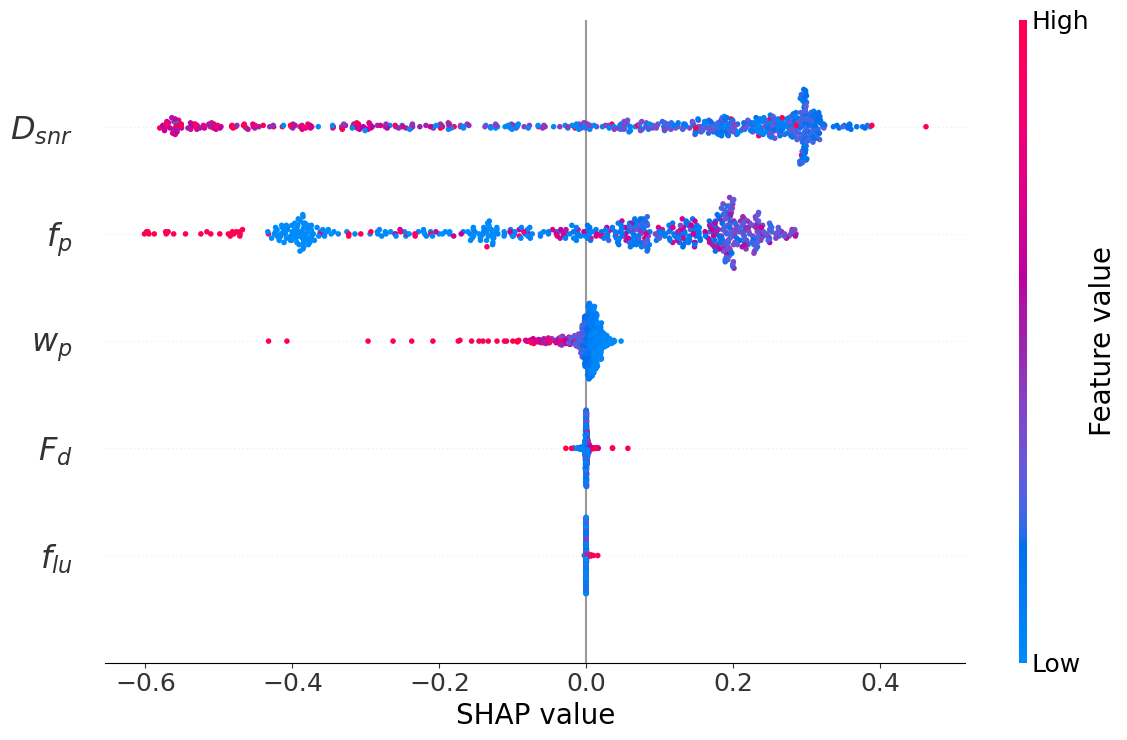}
    \end{subfigure}
    \begin{subfigure}{0.49\textwidth}
        \centering
        \texttt{LR}\\
        \includegraphics[width=\linewidth]{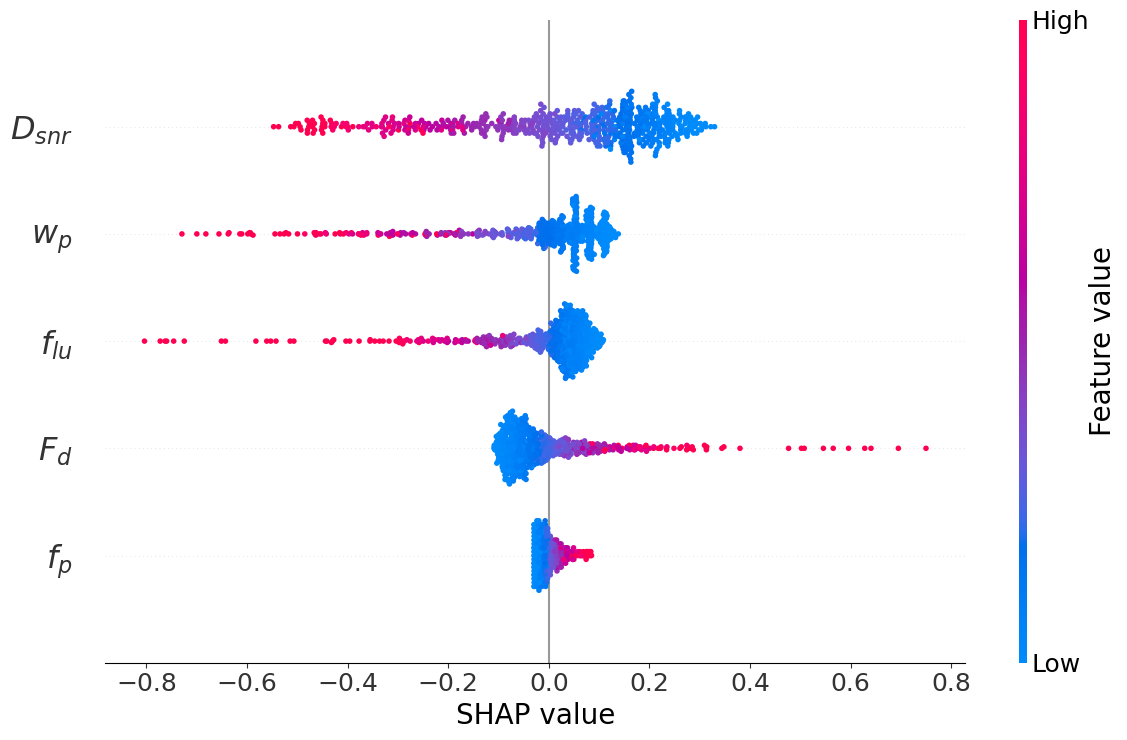}
    \end{subfigure}
    \begin{subfigure}{0.49\textwidth}
        \centering
        \texttt{Ada}\\
        \includegraphics[width=\linewidth]{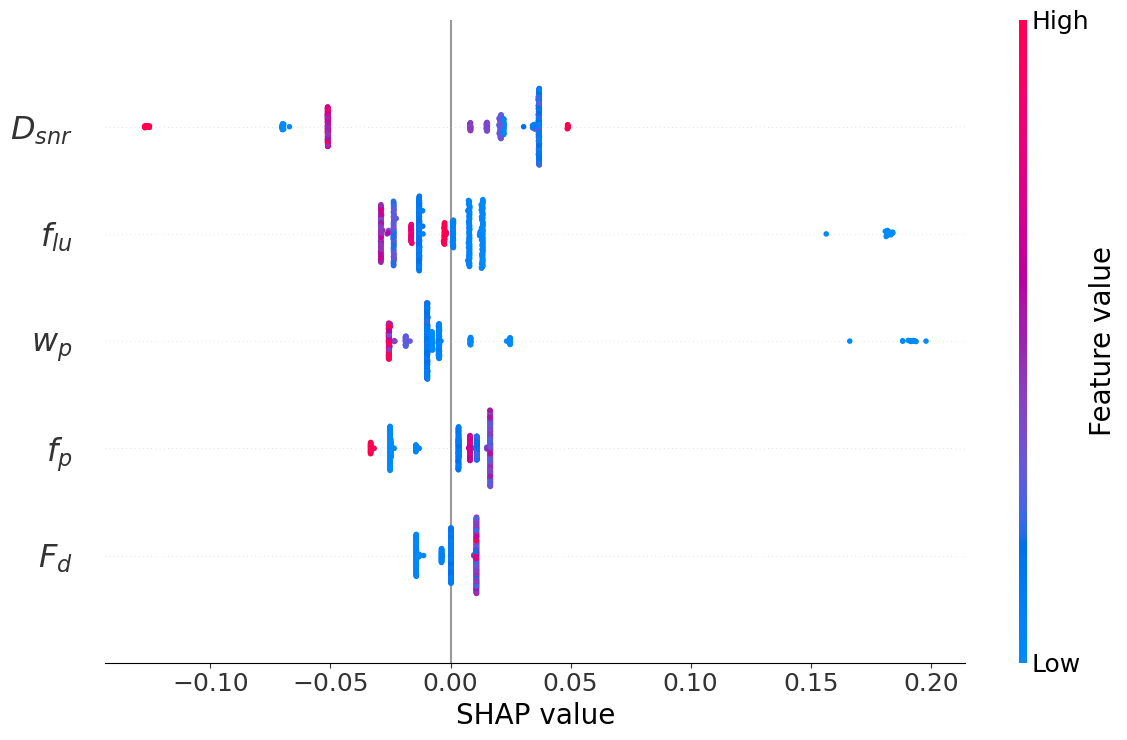}
    \end{subfigure}
    \begin{subfigure}{0.49\textwidth}
        \centering
        \texttt{GB}\\
        \includegraphics[width=\linewidth]{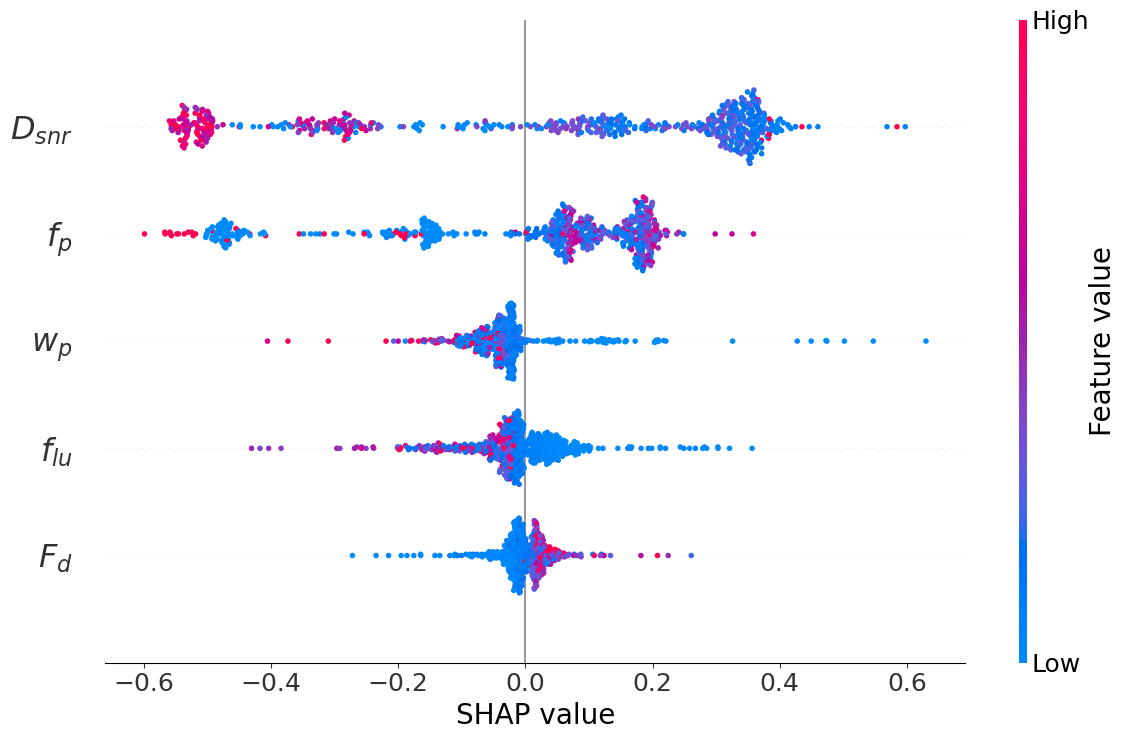}
    \end{subfigure}
    \caption{SHAP feature importance visualizations at the 1000th iteration for five different classifiers: RF, SVC, LR, Ada, and GB. Each point represents a data instance, colored by the feature value (low to high). The x-axis shows the SHAP value, which quantifies the impact of that feature on the model's prediction for a given sample. A high positive SHAP value indicates a strong contribution toward predicting a repeater (Class 1), while a large negative SHAP value reflects a strong influence toward predicting a non-repeater (Class 0).}
    \label{fig:five_models}
\end{figure*}

To better understand how each feature contributes to the model predictions, we examine SHAP values, which provide a unified measure of feature importance for individual samples. An example of SHAP feature importance at the 1000th iteration of the train-test split is shown in Fig.~\ref{fig:five_models}. In each subfigure, features are ordered from top to bottom by their relative importance, with the most important feature at the top. These plots provide insight into how each model utilizes the five input features to distinguish between repeater and non-repeater classes. Large positive and negative SHAP values indicate strong contributions toward predicting repeaters and non-repeaters, respectively. Across all base estimators, $D_{snr}$ is consistently identified as the most important feature, with high $D_{snr}$ values typically associated with non-repeaters and low $D_{snr}$ values associated with repeaters. However, it should be noted that the machine learning models do not rely solely on a single feature for prediction, and each base estimator may weigh features differently. For RF, SVC, and GB, the top three most important features are $D_{snr}$, $f_p$, and $w_p$; for LR and Ada, they are $D_{snr}$, $w_p$, and $F_d$.


\begin{figure*}
    \centering
    \begin{subfigure}{0.95\textwidth}
        \includegraphics[width=\linewidth]{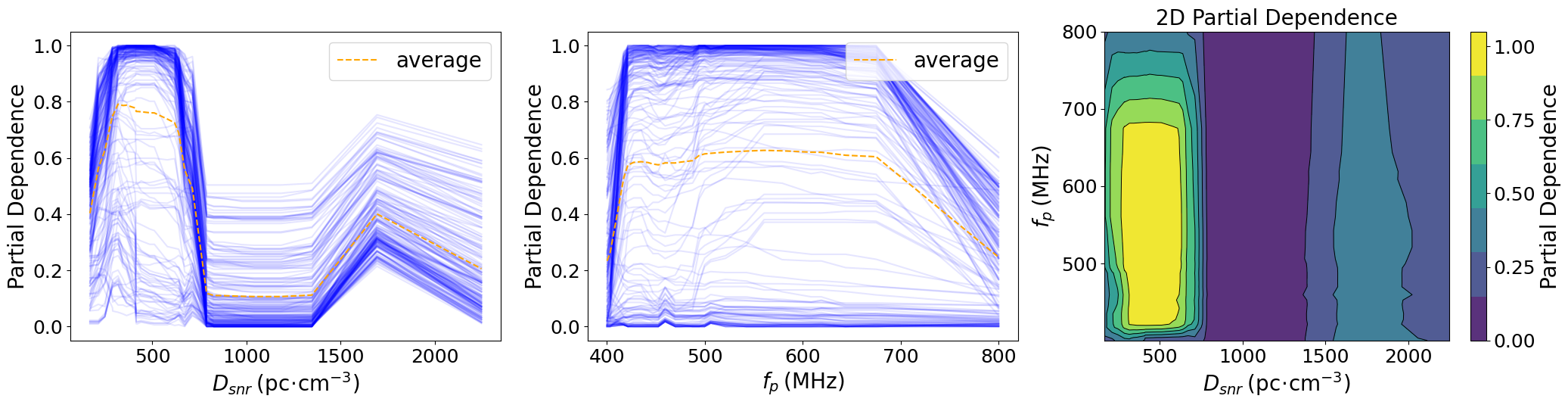}
    \end{subfigure}%
    \caption{Partial dependence plots for the most important feature ($D_{snr}$), along with the second most important features ($f_{p}$) in the RF model. The left and middle panels show the one-way partial dependence of each feature, while the right panel displays the two-way interaction between two features. High partial dependence values indicate that the feature values strongly influence the model toward predicting Class 1 (repeater), whereas low values suggest a tendency toward Class 0 (non-repeater). The blue lines represent the individual partial dependencies for each data point, while the orange dashed line denotes the average marginal effect of the model with respect to that feature. For the right panel, contour lines delineate regions of equal partial dependence. Yellow and green regions correspond to higher model-predicted probabilities of repeaters, whereas purple and blue regions indicate higher probabilities of non-repeaters.}
    \label{fig:pd}
\end{figure*}

Fig.~\ref{fig:pd} presents, as an example, partial dependence plots (PDPs) from the RF model for the most important feature, $D_{snr}$, together with the second-ranked features, $f_p$. The one-dimensional PDPs (left and middle panels) illustrate how the predicted probability of being a repeater (Class 1) varies as each feature changes, averaged over all other features. High partial dependence values (closer to 1) indicate a strong tendency toward predicting Class 1, whereas low values (closer to 0) suggest a tendency toward Class 0 (non-repeater). The two-dimensional PDPs (right panels) further reveal interaction effects between $D_{snr}$ and $f_p$. In these plots, yellow and green regions correspond to higher probabilities of repeaters, while purple and blue regions correspond to non-repeaters. The contours show clear thresholds: repeaters are most likely when $D_{snr}$ is relatively low ($\lesssim 600$~$\rm pc\cdot cm^{-3})$ combined with moderate $f_p$ of $\sim 450$--700 MHz, whereas parameters outside these ranges likely drive the prediction strongly toward non-repeater classification. Taken together with the SHAP analysis, these results provide complementary perspectives on feature contributions and decision boundaries. While the illustration is shown here for the RF model, it is intended as a representative example of how our machine learning models operate in practice.


\section{Discussion and Conclusion}


Overall, our self-training pipeline combines UMAP and a nearest-neighbor approach to identify non-repeater sources (Class 0) that may have the potential to be repeaters (Class 1), which are then treated as an unlabeled class (Class -1). A self-training machine learning framework is subsequently applied to classify these unlabeled and non-repeater sources, identifying repeater candidates (Class 2). Using this approach, we identify 168 repeater candidates with high model efficiency, achieving recall, precision, and $F_{1}$-scores above $\sim$83\%. The obtained accuracy is comparable to those of previous studies, yet our method relies on only five features rather than a larger set of features (e.g., \citet{luo2023, yang2023,  Sharma2024}), demonstrating both efficiency and adaptability of the proposed method.

As shown in Section 2, the Mann–Whitney U test indicates the significance of the differences between the five features of repeaters and non-repeaters. Among them, $D_{snr}$ shows the strongest statistical separation with the lowest p-value, followed by $w_{p}$ and $f_{p}$. This finding is consistent with the machine learning results, as illustrated in, e.g., Fig. \ref{fig:five_models}, where these three features emerge as the most important for distinguishing repeaters from non-repeaters. However, note that Table~\ref{tab:repeater_stats} reports standard statistical tests on the original dataset and does not account for hidden repeaters. While the test suggests $f_{lu}$ is not significant, some ML models (e.g., Ada) rank it second, indicating that traditional statistics applying directly to the dataset alone may be insufficient. Moreover, each ML model operates independently and produces its own ranking of important parameters, underscoring the importance of the voting process adopted in this work and also in, e.g., \citet{luo2023, Sharma2024}.

Furthermore, while $D_{snr}$ is identified as the most important feature, its effect is not straightforward. Some previous studies suggest that the host galaxy and its immediate environment may play a dominant role in shaping the dispersion measure \citep{yang2017, straal2020, orr2024}, and some generally regarded the it as moderately informative \citep{xu2023, Sharma2024}). Here, we find that higher $D_{snr}$ values are more likely associated with non-repeaters, while lower $D_{snr}$ values tend to correspond to repeaters. This trend may reflect a distance effect, where distant sources (with higher $D_{snr}$) are more often classified as non-repeaters, and nearby sources (with lower $D_{snr}$) as repeaters. This raises the possibility that the separation between repeaters and non-repeaters could be driven by distance and contributions from the intergalactic medium, the host galaxy, and the local environment of the FRBs. Therefore, $D_{snr}$ cannot be directly interpreted as a difference in the intrinsic physical properties of the sources, as it may also reflect environmental effects. A clearer interpretation will depend on improved knowledge of the source environments and the physical mechanisms of FRBs, which remain uncertain at present. Moreover, the ML model does not rely on a single feature but rather combines all five features in its prediction, as illustrated as an example in Fig. \ref{fig:pd}, right panel. Still, it is only capture pairwise effects and do not fully represent the complete decision process of the model, which jointly utilizes all five features.

\begin{figure*}
    \centering
    \includegraphics[width=35pc]{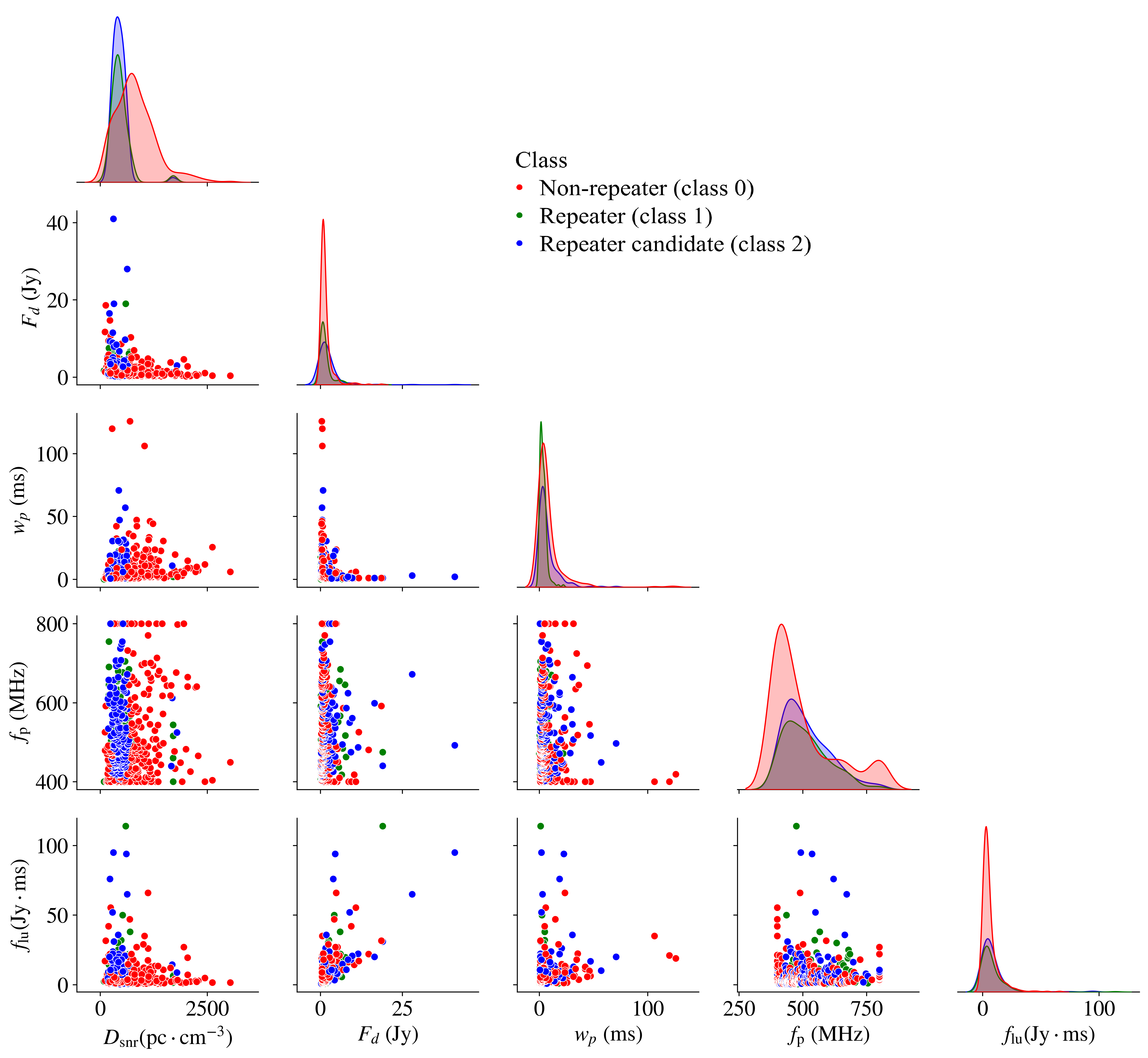}\\
    \caption{Pairplot illustrating the pairwise relationships and marginal distributions among the five extracted features. Note that Class 0, 1, and 2 are non-repeaters, known-repeaters, and repeater candidates, respectively.
    Notice that Class 2 (repeater candidates) exhibits a similar distribution to Class 1 (known repeaters). }
    \label{fig:pairplot_class123}
\end{figure*}

Given the complexity of the self-training pipeline, the feature importance of the machine learning model cannot be directly extracted. Instead, we examine the feature distributions of non-repeaters (Class 0), repeaters (Class 1), and repeater candidates (Class 2), as shown in Fig.~\ref{fig:pairplot_class123}. It can be seen that the feature distributions of repeaters and repeater candidates are strongly similar. To quantify this, we applied the Mann–Whitney U test to assess whether each feature differs significantly between Class 1 (repeaters) and Class 2 (candidates). The test results indicate that $D_{snr}$, $f_{p}$, and $f_{lu}$ do not differ significantly, with p-values greater than 0.05. This suggests that these features in repeaters and repeater candidates are likely drawn from the same population, thereby supporting the robustness of our method in identifying repeater candidates.

While our results suggests that $D_{snr}$, $f_{p}$, and $f_{lu}$ play a significant role in identifying repeater candidates, which is consistent with, e.g., \cite{chen2022, luo2023, xu2023, yang2023, zhu2023, Sharma2024}, we find that $w_{p}$ and $F_{d}$ have little impact on identifying repeater candidates. For $w_{p}$, previous studies report mixed findings: while some works \citep[e.g.][]{yang2023, Sharma2024} suggest it is an important discriminative feature, others are consistent with our conclusion that its role is limited \citep[e.g.][]{chen2022, luo2023, xu2023}. In contrast, for $F_{d}$, our result agree with the majority of studies that it is among the least informative features \citep[e.g.][]{luo2023, xu2023, yang2023, Sharma2024}.

\begin{figure}
    \centering
    \includegraphics[width=20pc]{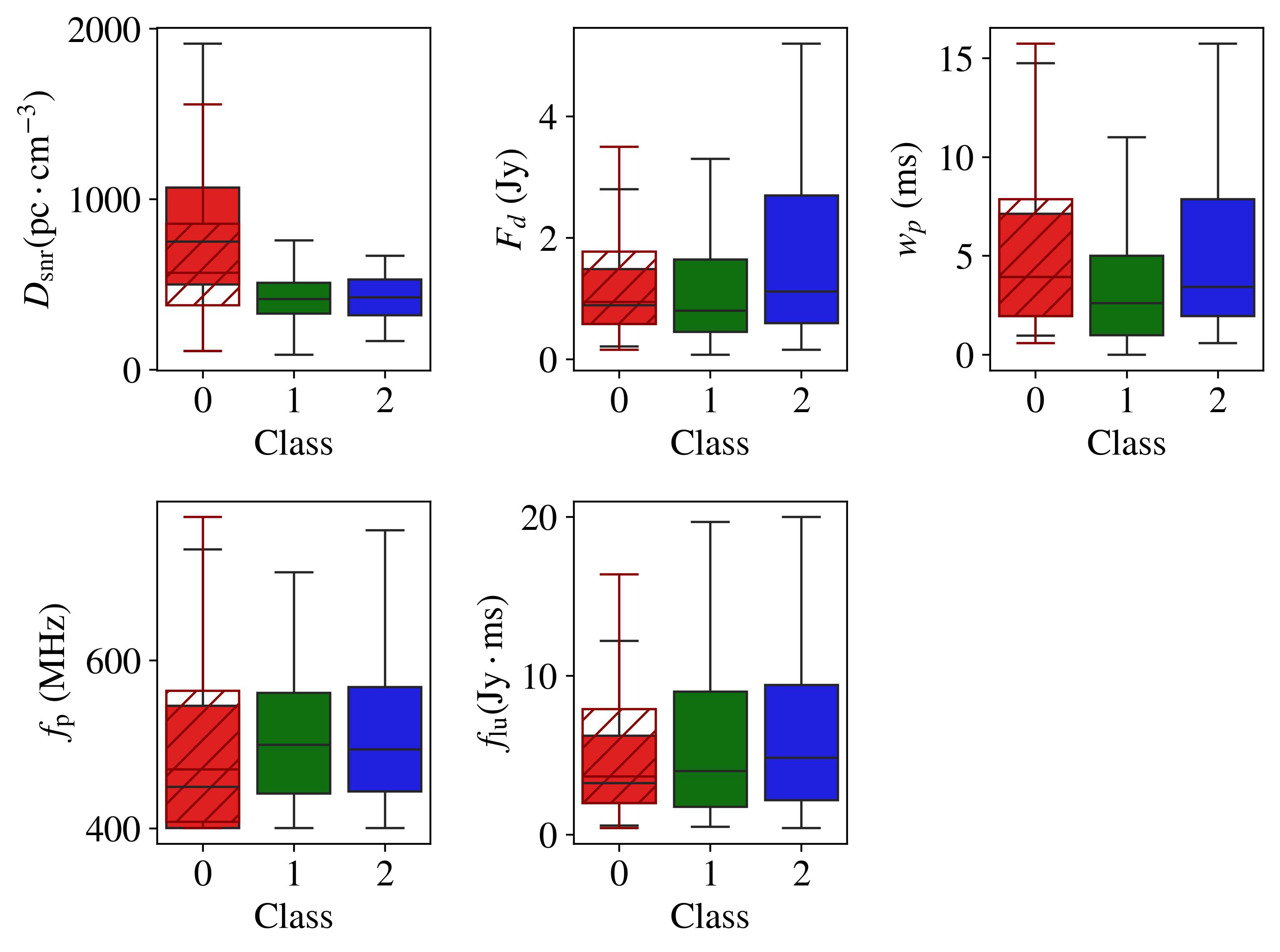}\\
    \caption{Box plots showing feature distributions for three FRB populations: confirmed repeaters (Class 1, solid green), updated non-repeaters (Class 0, solid red), and repeater candidates (Class 2, solid blue). Note that the full original non-repeater population is shown by the red hatching, and the repeater candidates are a subset of this population. The components of the box plots are defined as in Fig.~\ref{fig:boxplot}}
    \label{fig:boxplot_class02}
\end{figure}

Fig.~\ref{fig:boxplot_class02} presents box plots of the feature distributions for non-repeaters, repeaters, and repeater candidates. The red hatched corresponds to the distribution of non-repeaters (class 0) before some of them are reclassified as repeater candidates, which are shown in blue. Consistent with the Mann–Whitney U test results, the features $D_{snr}$, $f_{p}$, and $f_{lu}$ have p-values greater than 0.05, indicating that we cannot reject the null hypothesis that two independent groups come from the same population. This implies that repeaters (Class 1) and repeater candidates (Class 2) are statistically drawn from the same population. Specifically, the $D_{snr}$ distribution of non-repeaters shifts toward higher values, whereas both repeaters and repeater candidates exhibit lower $D_{snr}$ values. For $f_{p}$, non-repeaters are distributed at lower values, while repeaters and candidates show slightly higher values, in agreement with \cite{zhong2022} and \cite{Sharma2024}. For $f_{lu}$, non-repeaters are distributed at lower values, although this contradicts \cite{Sharma2024} which reported the opposite trend. Regarding $D_{snr}$, our findings are consistent with \cite{Sharma2024}, who reported that lower dispersion measure values are more predictive of repeaters, whereas higher dispersion measure values are more predictive of non-repeaters. \cite{curtin2024} found that non-repeaters tend to exhibit larger bandwidths associated with higher dispersion measures. This may imply that non-repeaters possess intrinsically higher specific energies than repeaters.

Many theories have been proposed regarding the origin of FRBs, and some have gained empirical support in recent years. \citet{bruni2024} provided strong observational confirmation of the plasma bubble engine model by examining FRB~20201124A. They found that the persistent radio emission follows expectations for an ionized nebular bubble around the central engine, directly linking its activity to the surrounding plasma. Such bubbles are thought to be driven by winds from a magnetar or a high-accretion X-ray binary, producing the observed persistent emission associated with the FRB. \cite{nimmo2025} offered the first direct evidence that FRBs can originate from neutron star magnetospheres. Through scintillation analysis of FRB~20221022A, they concluded that the burst formed within 10,000 km of a rotating neutron star, supporting the magnetospheric origin hypothesis. Meanwhile, \citet{eftekhari2025} reported FRB~20240209A in the outskirts of an old, quiescent elliptical galaxy, challenging both \citet{bruni2024} and \citet{nimmo2025} and implying alternative formation channels beyond young magnetar or core-collapse origins. If repeaters and non-repeaters arise from distinct progenitors or environments, the associated processes may leave environmental imprints on $D_{snr}$, highlighting its potential as a key discriminator between the two. However, \citet{james2023}, \citet{kirsten2024}, \citet{yamasaki2024} and \citet{Beniamini2025} suggested that repeaters and non-repeaters belong to a single population differing only by observability. Note that $D_{snr}$ represents the observed total DM and does not directly map to distance because the Milky Way contribution must first be subtracted. Nevertheless, population studies show that both $D_{snr}$ and the extragalactic DM differ statistically between repeaters and apparent non-repeaters \citep{chime2023}, indicating that $D_{snr}$ retains distance information. Then, our findings may also support the scenario in which repeaters tend to have lower $D_{snr}$, implying closer proximity and easier detection. Conversely, non-repeaters’ higher $D_{snr}$ may indicate greater distance, meaning they could burst again, but remain undetected given current sensitivity limits. In addition to the possibility that high-DM FRBs are intrinsically fainter, detectability decreases at large DM owing to intra-channel smearing. The First CHIME/FRB Catalog \citep{Amiri2021} shows that the DM selection function varies substantially, with some bias against high-DM events due to DM smearing. These instrumental effects, together with any intrinsic or distance-related trends, may explain the reduced number of high-DM repeater candidates recovered in our semi-supervised model.

In conclusion, our model suggests that repeaters are characterized by relatively lower average $D_{snr}$, higher average $f_{p}$, and higher $f_{lu}$ compared to non-repeaters, while no consistent differences are found in $w_{p}$ and $F_{d}$. Notably, the model achieves high accuracy with only a small set of features, underscoring the efficiency of our pipeline and showing that a minimal subset can capture the essential distinctions between repeaters and non-repeaters. A compact feature set reduces overfitting and improves interpretability, clarifying the physical implications of feature importance. We note, however, that CHIME’s 400--800 MHz frequency band may bias frequency-dependent features, and the limited number of labeled repeaters constrains robustness and generalizability. The result highlights the importance of $D_{snr}$ as a discrimator, suggesting a tension between physical and instrumental origins, either environmental effects or detection bias from nearby, easily observable sources. The ongoing accumulation of repeater detections, particularly with the release of CHIME/FRB Catalog 2, will provide larger labeled samples, enabling more rigorous validation of both feature importance and candidate identification.





\section*{Acknowledgements}
We thank the referee for the thoughtful suggestion that has helped improve the manuscript. This project is funded by National Research Council of Thailand (NRCT) and Suranaree University of Technology, grant number N42A680156. PC also thanks (i) Suranaree University of Technology (SUT), (ii) Thailand Science Research and Innovation (TSRI), and (iii) National Science, Research and Innovation Fund (Grant No. 204265). WL acknowledges financial support from the Faculty of Science, Srinakharinwirot University, through its internal funds for Fiscal Year 2025 (Grant No. 219/2568). We confirm that there are no conflicts of interest associated with this submission.

\section*{Data Availability}
The data and model analyzed in this article are available from the corresponding author upon reasonable request.



\bibliographystyle{mnras}
\bibliography{ref} 

@ARTICLE{cover1967,
       author = {{Cover}, T. and {Hart}, P.},
        title = "{Nearest neighbor pattern classification}",
      journal = {IEEE Transactions on Information Theory},
         year = 1967,
        month = jan,
       volume = {13},
       number = {1},
        pages = {21-27},
          doi = {10.1109/TIT.1967.1053964},
       adsurl = {https://ui.adsabs.harvard.edu/abs/1967ITIT...13...21C}
}

@article{friedman2001,
  title={Greedy function approximation: a gradient boosting machine},
  author={Friedman, Jerome H},
  journal={Annals of statistics},
  pages={1189--1232},
  year={2001},
  publisher={JSTOR}
}

@ARTICLE{chawla2002,
       author = {{Chawla}, N.~V. and {Bowyer}, K.~W. and {Hall}, L.~O. and {Kegelmeyer}, W.~P.},
        title = "{SMOTE: Synthetic Minority Over-sampling Technique}",
      journal = {arXiv e-prints},
         year = 2011,
        month = jun,
          eid = {arXiv:1106.1813},
        pages = {arXiv:1106.1813},
          doi = {10.48550/arXiv.1106.1813},
archivePrefix = {arXiv},
       eprint = {1106.1813},
 primaryClass = {cs.AI},
       adsurl = {https://ui.adsabs.harvard.edu/abs/2011arXiv1106.1813C}
}

@ARTICLE{Lorimer2007,
       author = {{Lorimer}, D.~R. and {Bailes}, M. and {McLaughlin}, M.~A. and {Narkevic}, D.~J. and {Crawford}, F.},
        title = "{A Bright Millisecond Radio Burst of Extragalactic Origin}",
      journal = {Science},
         year = 2007,
        month = nov,
       volume = {318},
       number = {5851},
        pages = {777},
          doi = {10.1126/science.1147532},
archivePrefix = {arXiv},
       eprint = {0709.4301},
 primaryClass = {astro-ph},
       adsurl = {https://ui.adsabs.harvard.edu/abs/2007Sci...318..777L}
}

@ARTICLE{scikit-learn,
       author = {{Pedregosa}, Fabian and {Varoquaux}, Ga{\"e}l and {Gramfort}, Alexandre and {Michel}, Vincent and {Thirion}, Bertrand and {Grisel}, Olivier and {Blondel}, Mathieu and {M{\"u}ller}, Andreas and {Nothman}, Joel and {Louppe}, Gilles and {Prettenhofer}, Peter and {Weiss}, Ron and {Dubourg}, Vincent and {Vanderplas}, Jake and {Passos}, Alexandre and {Cournapeau}, David and {Brucher}, Matthieu and {Perrot}, Matthieu and {Duchesnay}, {\'E}douard},
        title = "{Scikit-learn: Machine Learning in Python}",
      journal = {Journal of Machine Learning Research},
         year = 2011,
        month = oct,
       volume = {12},
        pages = {2825-2830},
          doi = {10.48550/arXiv.1201.0490},
archivePrefix = {arXiv},
       eprint = {1201.0490},
 primaryClass = {cs.LG},
       adsurl = {https://ui.adsabs.harvard.edu/abs/2011JMLR...12.2825P}
}

@misc{hastie2009,
  title={The elements of statistical learning},
  author={Hastie, Trevor and Tibshirani, Robert and Friedman, Jerome and others},
  year={2009},
  publisher={Springer series in statistics New-York}
}

@article{Thornton2013,
  author = {Thornton, D. and Stappers, B. and Bailes, M. and et al.},
  title = {A population of fast radio bursts at cosmological distances},
  journal = {Science},
  volume = {341},
  number = {6141},
  pages = {53--56},
  year = {2013},
  doi = {10.1126/science.1236789}
}

@article{Spitler2016,
  author = {Spitler, L. G. and Scholz, P. and Hessels, J. W. T. and et al.},
  title = {A repeating fast radio burst},
  journal = {Nature},
  volume = {531},
  number = {7593},
  pages = {202--205},
  year = {2016},
  doi = {10.1038/nature17168}
}

@article{Petroff2016,
  author = {Petroff, E. and Barr, E. D. and Jameson, A. and et al.},
  title = {FRBCAT: The Fast Radio Burst Catalogue},
  journal = {Publications of the Astronomical Society of Australia},
  volume = {33},
  pages = {e045},
  year = {2016},
  doi = {10.1017/pasa.2016.35}
}

@ARTICLE{Caleb2016,
       author = {{Caleb}, M. and {Flynn}, C. and {Bailes}, M. and {Barr}, E.~D. and {Hunstead}, R.~W. and {Keane}, E.~F. and {Ravi}, V. and {van Straten}, W.},
        title = "{Are the distributions of fast radio burst properties consistent with a cosmological population?}",
      journal = {\mnras},
         year = 2016,
        month = may,
       volume = {458},
       number = {1},
        pages = {708-717},
          doi = {10.1093/mnras/stw175},
archivePrefix = {arXiv},
       eprint = {1512.02738},
 primaryClass = {astro-ph.HE},
       adsurl = {https://ui.adsabs.harvard.edu/abs/2016MNRAS.458..708C}
}

@ARTICLE{lemaavztre2017,
       author = {{Lemaitre}, Guillaume and {Nogueira}, Fernando and {Aridas}, Christos K.},
        title = "{Imbalanced-learn: A Python Toolbox to Tackle the Curse of Imbalanced Datasets in Machine Learning}",
      journal = {arXiv e-prints},
         year = 2016,
        month = sep,
          eid = {arXiv:1609.06570},
        pages = {arXiv:1609.06570},
          doi = {10.48550/arXiv.1609.06570},
archivePrefix = {arXiv},
       eprint = {1609.06570},
 primaryClass = {cs.LG},
       adsurl = {https://ui.adsabs.harvard.edu/abs/2016arXiv160906570L}
}

@ARTICLE{lundberg2017,
       author = {{Lundberg}, Scott and {Lee}, Su-In},
        title = "{A Unified Approach to Interpreting Model Predictions}",
      journal = {arXiv e-prints},
         year = 2017,
        month = may,
          eid = {arXiv:1705.07874},
        pages = {arXiv:1705.07874},
          doi = {10.48550/arXiv.1705.07874},
archivePrefix = {arXiv},
       eprint = {1705.07874},
 primaryClass = {cs.AI},
       adsurl = {https://ui.adsabs.harvard.edu/abs/2017arXiv170507874L}
}

@ARTICLE{yang2017,
       author = {{Yang}, Yuan-Pei and {Luo}, Rui and {Li}, Zhuo and {Zhang}, Bing},
        title = "{Large Host-galaxy Dispersion Measure of Fast Radio Bursts}",
      journal = {\apjl},
         year = 2017,
        month = apr,
       volume = {839},
       number = {2},
          eid = {L25},
        pages = {L25},
          doi = {10.3847/2041-8213/aa6c2e},
archivePrefix = {arXiv},
       eprint = {1701.06465},
 primaryClass = {astro-ph.HE},
       adsurl = {https://ui.adsabs.harvard.edu/abs/2017ApJ...839L..25Y}
}

@ARTICLE{Bhandari2018,
       author = {{Bhandari}, S. and {Keane}, E.~F. and {Barr}, E.~D. and {Jameson}, A. and {Petroff}, E. and {Johnston}, S. and {Bailes}, M. and {Bhat}, N.~D.~R. and {Burgay}, M. and {Burke-Spolaor}, S. and {Caleb}, M. and {Eatough}, R.~P. and {Flynn}, C. and {Green}, J.~A. and {Jankowski}, F. and {Kramer}, M. and {Krishnan}, V. Venkatraman and {Morello}, V. and {Possenti}, A. and {Stappers}, B. and {Tiburzi}, C. and {van Straten}, W. and {Andreoni}, I. and {Butterley}, T. and {Chandra}, P. and {Cooke}, J. and {Corongiu}, A. and {Coward}, D.~M. and {Dhillon}, V.~S. and {Dodson}, R. and {Hardy}, L.~K. and {Howell}, E.~J. and {Jaroenjittichai}, P. and {Klotz}, A. and {Littlefair}, S.~P. and {Marsh}, T.~R. and {Mickaliger}, M. and {Muxlow}, T. and {Perrodin}, D. and {Pritchard}, T. and {Sawangwit}, U. and {Terai}, T. and {Tominaga}, N. and {Torne}, P. and {Totani}, T. and {Trois}, A. and {Turpin}, D. and {Niino}, Y. and {Wilson}, R.~W. and {Albert}, A. and {Andr{\'e}}, M. and {Anghinolfi}, M. and {Anton}, G. and {Ardid}, M. and {Aubert}, J.-J. and {Avgitas}, T. and {Baret}, B. and {Barrios-Mart{\'\i}}, J. and {Basa}, S. and {Belhorma}, B. and {Bertin}, V. and {Biagi}, S. and {Bormuth}, R. and {Bourret}, S. and {Bouwhuis}, M.~C. and {Br{\^a}nza{\c{s}}}, H. and {Bruijn}, R. and {Brunner}, J. and {Busto}, J. and {Capone}, A. and {Caramete}, L. and {Carr}, J. and {Celli}, S. and {Moursli}, R. Cherkaoui El and {Chiarusi}, T. and {Circella}, M. and {Coelho}, J.~A.~B. and {Coleiro}, A. and {Coniglione}, R. and {Costantini}, H. and {Coyle}, P. and {Creusot}, A. and {D{\'\i}az}, A.~F. and {Deschamps}, A. and {De Bonis}, G. and {Distefano}, C. and {Palma}, I. Di and {Domi}, A. and {Donzaud}, C. and {Dornic}, D. and {Drouhin}, D. and {Eberl}, T. and {Bojaddaini}, I. El and {Khayati}, N. El and {Els{\"a}sser}, D. and {Enzenh{\"o}fer}, A. and {Ettahiri}, A. and {Fassi}, F. and {Felis}, I. and {Fusco}, L.~A. and {Gay}, P. and {Giordano}, V. and {Glotin}, H. and {Gregoire}, T. and {Gracia-Ruiz}, R. and {Graf}, K. and {Hallmann}, S. and {van Haren}, H. and {Heijboer}, A.~J. and {Hello}, Y. and {Hern{\'a}ndez-Rey}, J.~J. and {H{\"o}{\ss}l}, J. and {Hofest{\"a}dt}, J. and {Hugon}, C. and {Illuminati}, G. and {James}, C.~W. and {de Jong}, M. and {Jongen}, M. and {Kadler}, M. and {Kalekin}, O. and {Katz}, U. and {Kie{\ss}ling}, D. and {Kouchner}, A. and {Kreter}, M. and {Kreykenbohm}, I. and {Kulikovskiy}, V. and {Lachaud}, C. and {Lahmann}, R. and {Lef{\`e}vre}, D. and {Leonora}, E. and {Loucatos}, S. and {Marcelin}, M. and {Margiotta}, A. and {Marinelli}, A. and {Mart{\'\i}nez-Mora}, J.~A. and {Mele}, R. and {Melis}, K. and {Michael}, T. and {Migliozzi}, P. and {Moussa}, A. and {Navas}, S. and {Nezri}, E. and {Organokov}, M. and {P{\v{a}}v{\v{a}}la{\c{s}}}, G.~E. and {Pellegrino}, C. and {Perrina}, C. and {Piattelli}, P. and {Popa}, V. and {Pradier}, T. and {Quinn}, L. and {Racca}, C. and {Riccobene}, G. and {S{\'a}nchez-Losa}, A. and {Salda{\~n}a}, M. and {Salvadori}, I. and {Samtleben}, D.~F.~E. and {Sanguineti}, M. and {Sapienza}, P. and {Sch{\"u}ssler}, F. and {Sieger}, C. and {Spurio}, M. and {Stolarczyk}, Th and {Taiuti}, M. and {Tayalati}, Y. and {Trovato}, A. and {Turpin}, D. and {T{\"o}nnis}, C. and {Vallage}, B. and {Van Elewyck}, V. and {Versari}, F. and {Vivolo}, D. and {Vizzocca}, A. and {Wilms}, J. and {Zornoza}, J.~D. and {Z{\'u}{\~n}iga}, J.},
        title = "{The SUrvey for Pulsars and Extragalactic Radio Bursts - II. New FRB discoveries and their follow-up}",
      journal = {\mnras},
         year = 2018,
        month = apr,
       volume = {475},
       number = {2},
        pages = {1427-1446},
          doi = {10.1093/mnras/stx3074},
archivePrefix = {arXiv},
       eprint = {1711.08110},
 primaryClass = {astro-ph.HE},
       adsurl = {https://ui.adsabs.harvard.edu/abs/2018MNRAS.475.1427B}
}

@ARTICLE{mcinnes2018,
       author = {{McInnes}, Leland and {Healy}, John and {Melville}, James},
        title = "{UMAP: Uniform Manifold Approximation and Projection for Dimension Reduction}",
      journal = {arXiv e-prints},
         year = 2018,
        month = feb,
          eid = {arXiv:1802.03426},
        pages = {arXiv:1802.03426},
          doi = {10.48550/arXiv.1802.03426},
archivePrefix = {arXiv},
       eprint = {1802.03426},
 primaryClass = {stat.ML},
       adsurl = {https://ui.adsabs.harvard.edu/abs/2018arXiv180203426M}
}

@article{Shannon2018,
  author = {Shannon, R. M. and Macquart, J. P. and Bannister, K. W. and et al.},
  title = {The dispersion-brightness relation for fast radio bursts from a wide-field survey},
  journal = {Nature},
  volume = {562},
  number = {7727},
  pages = {386--390},
  year = {2018},
  doi = {10.1038/s41586-018-0588-y}
}

@ARTICLE{Petroff2019,
       author = {{Petroff}, E. and {Hessels}, J.~W.~T. and {Lorimer}, D.~R.},
        title = "{Fast radio bursts}",
      journal = {\aapr},
         year = 2019,
        month = dec,
       volume = {27},
       number = {1},
          eid = {4},
        pages = {4},
          doi = {10.1007/s00159-019-0116-6},
archivePrefix = {arXiv},
       eprint = {1904.07947},
 primaryClass = {astro-ph.HE},
       adsurl = {https://ui.adsabs.harvard.edu/abs/2019A&ARv..27....4P}
}

@ARTICLE{CHIME2019,
       author = {{CHIME/FRB Collaboration} and {Amiri}, M. and {Bandura}, K. and {Bhardwaj}, M. and {Boubel}, P. and {Boyce}, M.~M. and {Boyle}, P. J and {. Brar}, C. and {Burhanpurkar}, M. and {Cassanelli}, T. and {Chawla}, P. and {Cliche}, J.~F. and {Cubranic}, D. and {Deng}, M. and {Denman}, N. and {Dobbs}, M. and {Fandino}, M. and {Fonseca}, E. and {Gaensler}, B.~M. and {Gilbert}, A.~J. and {Gill}, A. and {Giri}, U. and {Good}, D.~C. and {Halpern}, M. and {Hanna}, D.~S. and {Hill}, A.~S. and {Hinshaw}, G. and {H{\"o}fer}, C. and {Josephy}, A. and {Kaspi}, V.~M. and {Landecker}, T.~L. and {Lang}, D.~A. and {Lin}, H.-H. and {Masui}, K.~W. and {Mckinven}, R. and {Mena-Parra}, J. and {Merryfield}, M. and {Michilli}, D. and {Milutinovic}, N. and {Moatti}, C. and {Naidu}, A. and {Newburgh}, L.~B. and {Ng}, C. and {Patel}, C. and {Pen}, U. and {Pinsonneault-Marotte}, T. and {Pleunis}, Z. and {Rafiei-Ravandi}, M. and {Rahman}, M. and {Ransom}, S.~M. and {Renard}, A. and {Scholz}, P. and {Shaw}, J.~R. and {Siegel}, S.~R. and {Smith}, K.~M. and {Stairs}, I.~H. and {Tendulkar}, S.~P. and {Tretyakov}, I. and {Vanderlinde}, K. and {Yadav}, P.},
        title = "{A second source of repeating fast radio bursts}",
      journal = {\nat},
         year = 2019,
        month = jan,
       volume = {566},
       number = {7743},
        pages = {235-238},
          doi = {10.1038/s41586-018-0864-x},
archivePrefix = {arXiv},
       eprint = {1901.04525},
 primaryClass = {astro-ph.HE},
       adsurl = {https://ui.adsabs.harvard.edu/abs/2019Natur.566..235C}
}

@article{Heiser2020,
title = {A Quantitative Framework for Evaluating Single-Cell Data Structure Preservation by Dimensionality Reduction Techniques},
journal = {Cell Reports},
volume = {31},
number = {5},
pages = {107576},
year = {2020},
issn = {2211-1247},
doi = {https://doi.org/10.1016/j.celrep.2020.107576},
url = {https://www.sciencedirect.com/science/article/pii/S2211124720305258},
author = {Cody N. Heiser and Ken S. Lau}
}

@ARTICLE{connor2020,
       author = {{Connor}, L. and {Miller}, M.~C. and {Gardenier}, D.~W.},
        title = "{Beaming as an explanation of the repetition/width relation in FRBs}",
      journal = {\mnras},
         year = 2020,
        month = sep,
       volume = {497},
       number = {3},
        pages = {3076-3082},
          doi = {10.1093/mnras/staa2074},
archivePrefix = {arXiv},
       eprint = {2003.11930},
 primaryClass = {astro-ph.HE},
       adsurl = {https://ui.adsabs.harvard.edu/abs/2020MNRAS.497.3076C}
}

@article{CHIME2020,
  author = {{CHIME/FRB Collaboration}},
  title = {Periodic activity from a fast radio burst source},
  journal = {Nature},
  volume = {582},
  number = {7812},
  pages = {351--355},
  year = {2020},
  doi = {10.1038/s41586-020-2398-2}
}

@article{lundberg2020,
  title={From local explanations to global understanding with explainable AI for trees},
  author={Lundberg, Scott M. and Erion, Gabriel and Chen, Hugh and DeGrave, Alex and Prutkin, Jordan M. and Nair, Bala and Katz, Ronit and Himmelfarb, Jonathan and Bansal, Nisha and Lee, Su-In},
  journal={Nature Machine Intelligence},
  volume={2},
  number={1},
  pages={2522-5839},
  year={2020},
  publisher={Nature Publishing Group}
}

@ARTICLE{straal2020,
       author = {{Straal}, S.~M. and {Connor}, L. and {van Leeuwen}, J.},
        title = "{A dispersion excess from pulsar wind nebulae and supernova remnants: Implications for pulsars and FRBs}",
      journal = {\aap},
         year = 2020,
        month = feb,
       volume = {634},
          eid = {A105},
        pages = {A105},
          doi = {10.1051/0004-6361/201833376},
archivePrefix = {arXiv},
       eprint = {2001.06019},
 primaryClass = {astro-ph.HE},
       adsurl = {https://ui.adsabs.harvard.edu/abs/2020A&A...634A.105S}
}

@ARTICLE{zhang2020,
       author = {{Zhang}, Bing},
        title = "{The physical mechanisms of fast radio bursts}",
      journal = {\nat},
         year = 2020,
        month = nov,
       volume = {587},
       number = {7832},
        pages = {45-53},
          doi = {10.1038/s41586-020-2828-1},
archivePrefix = {arXiv},
       eprint = {2011.03500},
 primaryClass = {astro-ph.HE},
       adsurl = {https://ui.adsabs.harvard.edu/abs/2020Natur.587...45Z}
}

@ARTICLE{Amiri2021,
       author = {{CHIME/FRB Collaboration} and {Amiri}, Mandana and {Andersen}, Bridget C. and {Bandura}, Kevin and {Berger}, Sabrina and {Bhardwaj}, Mohit and {Boyce}, Michelle M. and {Boyle}, P.~J. and {Brar}, Charanjot and {Breitman}, Daniela and {Cassanelli}, Tomas and {Chawla}, Pragya and {Chen}, Tianyue and {Cliche}, J.-F. and {Cook}, Amanda and {Cubranic}, Davor and {Curtin}, Alice P. and {Deng}, Meiling and {Dobbs}, Matt and {Dong}, Fengqiu Adam and {Eadie}, Gwendolyn and {Fandino}, Mateus and {Fonseca}, Emmanuel and {Gaensler}, B.~M. and {Giri}, Utkarsh and {Good}, Deborah C. and {Halpern}, Mark and {Hill}, Alex S. and {Hinshaw}, Gary and {Josephy}, Alexander and {Kaczmarek}, Jane F. and {Kader}, Zarif and {Kania}, Joseph W. and {Kaspi}, Victoria M. and {Landecker}, T.~L. and {Lang}, Dustin and {Leung}, Calvin and {Li}, Dongzi and {Lin}, Hsiu-Hsien and {Masui}, Kiyoshi W. and {McKinven}, Ryan and {Mena-Parra}, Juan and {Merryfield}, Marcus and {Meyers}, Bradley W. and {Michilli}, Daniele and {Milutinovic}, Nikola and {Mirhosseini}, Arash and {M{\"u}nchmeyer}, Moritz and {Naidu}, Arun and {Newburgh}, Laura and {Ng}, Cherry and {Patel}, Chitrang and {Pen}, Ue-Li and {Petroff}, Emily and {Pinsonneault-Marotte}, Tristan and {Pleunis}, Ziggy and {Rafiei-Ravandi}, Masoud and {Rahman}, Mubdi and {Ransom}, Scott M. and {Renard}, Andre and {Sanghavi}, Pranav and {Scholz}, Paul and {Shaw}, J. Richard and {Shin}, Kaitlyn and {Siegel}, Seth R. and {Sikora}, Andrew E. and {Singh}, Saurabh and {Smith}, Kendrick M. and {Stairs}, Ingrid and {Tan}, Chia Min and {Tendulkar}, S.~P. and {Vanderlinde}, Keith and {Wang}, Haochen and {Wulf}, Dallas and {Zwaniga}, A.~V.},
        title = "{The First CHIME/FRB Fast Radio Burst Catalog}",
      journal = {\apjs},
         year = 2021,
        month = dec,
       volume = {257},
       number = {2},
          eid = {59},
        pages = {59},
          doi = {10.3847/1538-4365/ac33ab},
archivePrefix = {arXiv},
       eprint = {2106.04352},
 primaryClass = {astro-ph.HE},
       adsurl = {https://ui.adsabs.harvard.edu/abs/2021ApJS..257...59C}
}

@article{Li2021,
  author = {Li, D. and Wang, P. and Zhu, W. W. and et al.},
  title = {A bimodal burst energy distribution of a repeating fast radio burst source},
  journal = {Nature},
  volume = {598},
  number = {7882},
  pages = {267--271},
  year = {2021},
  doi = {10.1038/s41586-021-03878-5}
}

@software{head2021,
       author = {{Head}, Tim and {Kumar}, Manoj and {Nahrstaedt}, Holger and {Louppe}, Gilles and {Shcherbatyi}, Iaroslav},
        title = "{scikit-optimize/scikit-optimize}",
         year = 2021,
        month = oct,
          eid = {10.5281/zenodo.5565057},
          doi = {10.5281/zenodo.5565057},
      version = {v0.9.0},
    publisher = {Zenodo},
       adsurl = {https://ui.adsabs.harvard.edu/abs/2021zndo...5565057H}
}

@ARTICLE{chen2022,
       author = {{Chen}, Bo Han and {Hashimoto}, Tetsuya and {Goto}, Tomotsugu and {Kim}, Seong Jin and {Santos}, Daryl Joe D. and {On}, Alvina Y.~L. and {Lu}, Ting-Yi and {Hsiao}, Tiger Y.-Y.},
        title = "{Uncloaking hidden repeating fast radio bursts with unsupervised machine learning}",
      journal = {\mnras},
         year = 2022,
        month = jan,
       volume = {509},
       number = {1},
        pages = {1227-1236},
          doi = {10.1093/mnras/stab2994},
archivePrefix = {arXiv},
       eprint = {2110.09440},
 primaryClass = {astro-ph.HE},
       adsurl = {https://ui.adsabs.harvard.edu/abs/2022MNRAS.509.1227C}
}

@ARTICLE{zhong2022,
       author = {{Zhong}, Shu-Qing and {Xie}, Wen-Jin and {Deng}, Can-Min and {Li}, Long and {Dai}, Zi-Gao and {Zhang}, Hai-Ming},
        title = "{Can a Single Population Account for the Discriminant Properties in Fast Radio Bursts?}",
      journal = {\apj},
         year = 2022,
        month = feb,
       volume = {926},
       number = {2},
          eid = {206},
        pages = {206},
          doi = {10.3847/1538-4357/ac4d98},
archivePrefix = {arXiv},
       eprint = {2202.04422},
 primaryClass = {astro-ph.HE},
       adsurl = {https://ui.adsabs.harvard.edu/abs/2022ApJ...926..206Z}
}

@ARTICLE{yang2023,
       author = {{Yang}, X. and {Zhang}, S.-B. and {Wang}, J.-S. and {Wu}, X.-F.},
        title = "{Classifying FRB spectrograms using nonlinear dimensionality reduction techniques}",
      journal = {\mnras},
         year = 2023,
        month = jul,
       volume = {522},
       number = {3},
        pages = {4342-4351},
          doi = {10.1093/mnras/stad1304},
archivePrefix = {arXiv},
       eprint = {2304.13912},
 primaryClass = {astro-ph.HE},
       adsurl = {https://ui.adsabs.harvard.edu/abs/2023MNRAS.522.4342Y}
}

@ARTICLE{xu2023,
       author = {{Xu}, Jiaying and {Feng}, Yi and {Li}, Di and {Wang}, Pei and {Zhang}, Yongkun and {Xie}, Jintao and {Chen}, Huaxi and {Wang}, Han and {Kang}, Zhixuan and {Hu}, Jingjing and {Zheng}, Yun and {Tsai}, Chao-Wei and {Chen}, Xianglei and {Zhou}, Dengke},
        title = "{Blinkverse: A Database of Fast Radio Bursts}",
      journal = {Universe},
         year = 2023,
        month = jul,
       volume = {9},
       number = {7},
          eid = {330},
        pages = {330},
          doi = {10.3390/universe9070330},
archivePrefix = {arXiv},
       eprint = {2308.00336},
 primaryClass = {astro-ph.HE},
       adsurl = {https://ui.adsabs.harvard.edu/abs/2023Univ....9..330X}
}

@ARTICLE{luo2023,
       author = {{Luo}, Jia-Wei and {Zhu-Ge}, Jia-Ming and {Zhang}, Bing},
        title = "{Machine learning classification of CHIME fast radio bursts - I. Supervised methods}",
      journal = {\mnras},
         year = 2023,
        month = jan,
       volume = {518},
       number = {2},
        pages = {1629-1641},
          doi = {10.1093/mnras/stac3206},
archivePrefix = {arXiv},
       eprint = {2210.02463},
 primaryClass = {astro-ph.HE},
       adsurl = {https://ui.adsabs.harvard.edu/abs/2023MNRAS.518.1629L}
}

@ARTICLE{zhu2023,
       author = {{Zhu-Ge}, Jia-Ming and {Luo}, Jia-Wei and {Zhang}, Bing},
        title = "{Machine learning classification of CHIME fast radio bursts - II. Unsupervised methods}",
      journal = {\mnras},
         year = 2023,
        month = feb,
       volume = {519},
       number = {2},
        pages = {1823-1836},
          doi = {10.1093/mnras/stac3599},
archivePrefix = {arXiv},
       eprint = {2210.02471},
 primaryClass = {astro-ph.HE},
       adsurl = {https://ui.adsabs.harvard.edu/abs/2023MNRAS.519.1823Z}
}

@ARTICLE{james2023,
       author = {{James}, C.~W.},
        title = "{Modelling repetition in zDM: A single population of repeating fast radio bursts can explain CHIME data}",
      journal = {\pasa},
         year = 2023,
        month = dec,
       volume = {40},
          eid = {e057},
        pages = {e057},
          doi = {10.1017/pasa.2023.51},
archivePrefix = {arXiv},
       eprint = {2306.17403},
 primaryClass = {astro-ph.HE},
       adsurl = {https://ui.adsabs.harvard.edu/abs/2023PASA...40...57J}
}

@ARTICLE{chime2023,
       author = {{CHIME/FRB Collaboration} and {Andersen}, Bridget C. and {Bandura}, Kevin and {Bhardwaj}, Mohit and {Boyle}, P.~J. and {Brar}, Charanjot and {Cassanelli}, Tomas and {Chatterjee}, S. and {Chawla}, Pragya and {Cook}, Amanda M. and {Curtin}, Alice P. and {Dobbs}, Matt and {Dong}, Fengqiu Adam and {Faber}, Jakob T. and {Fandino}, Mateus and {Fonseca}, Emmanuel and {Gaensler}, B.~M. and {Giri}, Utkarsh and {Herrera-Martin}, Antonio and {Hill}, Alex S. and {Ibik}, Adaeze and {Josephy}, Alexander and {Kaczmarek}, Jane F. and {Kader}, Zarif and {Kaspi}, Victoria and {Landecker}, T.~L. and {Lanman}, Adam E. and {Lazda}, Mattias and {Leung}, Calvin and {Lin}, Hsiu-Hsien and {Masui}, Kiyoshi W. and {McKinven}, Ryan and {Mena-Parra}, Juan and {Meyers}, Bradley W. and {Michilli}, D. and {Ng}, Cherry and {Pandhi}, Ayush and {Pearlman}, Aaron B. and {Pen}, Ue-Li and {Petroff}, Emily and {Pleunis}, Ziggy and {Rafiei-Ravandi}, Masoud and {Rahman}, Mubdi and {Ransom}, Scott M. and {Renard}, Andre and {Sand}, Ketan R. and {Sanghavi}, Pranav and {Scholz}, Paul and {Shah}, Vishwangi and {Shin}, Kaitlyn and {Siegel}, Seth and {Smith}, Kendrick and {Stairs}, Ingrid and {Su}, Jianing and {Tendulkar}, Shriharsh P. and {Vanderlinde}, Keith and {Wang}, Haochen and {Wulf}, Dallas and {Zwaniga}, Andrew},
        title = "{CHIME/FRB Discovery of 25 Repeating Fast Radio Burst Sources}",
      journal = {\apj},
         year = 2023,
        month = apr,
       volume = {947},
       number = {2},
          eid = {83},
        pages = {83},
          doi = {10.3847/1538-4357/acc6c1},
archivePrefix = {arXiv},
       eprint = {2301.08762},
 primaryClass = {astro-ph.HE},
       adsurl = {https://ui.adsabs.harvard.edu/abs/2023ApJ...947...83C}
}

@unknown{Brown2023,
author = {Brown, Joseph and Mohapatra, Somesh and Lee, Michael and Misteli, Roman and Tseo, Yitong and Grob, Nathalie and Quartararo, Anthony and Loas, Andrei and Gomez-Bombarelli, Rafael and Pentelute, Bradley},
year = {2023},
month = {05},
pages = {},
title = {Unsupervised machine learning leads to an abiotic picomolar peptide ligand},
archivePrefix = {ChemRxiv},
doi = {10.26434/chemrxiv-2023-tws4n},
}

@ARTICLE{sharma2024,
       author = {{Sharma}, Arjun and {Rajpaul}, Vinesh Maguire},
        title = "{Positive and unlabelled machine learning reveals new fast radio burst repeater candidates}",
      journal = {\mnras},
         year = 2024,
        month = sep,
       volume = {533},
       number = {3},
        pages = {3283-3295},
          doi = {10.1093/mnras/stae1972},
archivePrefix = {arXiv},
       eprint = {2408.11436},
 primaryClass = {astro-ph.HE},
       adsurl = {https://ui.adsabs.harvard.edu/abs/2024MNRAS.533.3283S}
}

@ARTICLE{orr2024,
       author = {{Orr}, Matthew E. and {Burkhart}, Blakesley and {Lu}, Wenbin and {Ponnada}, Sam B. and {Hummels}, Cameron B.},
        title = "{Objects May Be Closer than They Appear: Significant Host Galaxy Dispersion Measures of Fast Radio Bursts in Zoom-in Simulations}",
      journal = {\apjl},
         year = 2024,
        month = sep,
       volume = {972},
       number = {2},
          eid = {L26},
        pages = {L26},
          doi = {10.3847/2041-8213/ad725b},
archivePrefix = {arXiv},
       eprint = {2406.03523},
 primaryClass = {astro-ph.GA},
       adsurl = {https://ui.adsabs.harvard.edu/abs/2024ApJ...972L..26O}
}

@ARTICLE{qiang2025,
       author = {{Qiang}, Da-Chun and {Zheng}, Jie and {You}, Zhi-Qiang and {Yang}, Sheng},
        title = "{Unsupervised Machine Learning for Classifying CHIME Fast Radio Bursts and Investigating Empirical Relations}",
      journal = {\apj},
         year = 2025,
        month = mar,
       volume = {982},
       number = {1},
          eid = {16},
        pages = {16},
          doi = {10.3847/1538-4357/adb72b},
archivePrefix = {arXiv},
       eprint = {2411.14040},
 primaryClass = {astro-ph.HE},
       adsurl = {https://ui.adsabs.harvard.edu/abs/2025ApJ...982...16Q}
}

@ARTICLE{bruni2024,
       author = {{Bruni}, Gabriele and {Piro}, Luigi and {Yang}, Yuan-Pei and {Quai}, Salvatore and {Zhang}, Bing and {Palazzi}, Eliana and {Nicastro}, Luciano and {Feruglio}, Chiara and {Tripodi}, Roberta and {O'Connor}, Brendan and {Gardini}, Angela and {Savaglio}, Sandra and {Rossi}, Andrea and {Nicuesa Guelbenzu}, Ana M. and {Paladino}, Rosita},
        title = "{A nebular origin for the persistent radio emission of fast radio bursts}",
      journal = {\nat},
         year = 2024,
        month = aug,
       volume = {632},
       number = {8027},
        pages = {1014-1016},
          doi = {10.1038/s41586-024-07782-6},
archivePrefix = {arXiv},
       eprint = {2312.15296},
 primaryClass = {astro-ph.HE},
       adsurl = {https://ui.adsabs.harvard.edu/abs/2024Natur.632.1014B}
}

@ARTICLE{kirsten2024,
       author = {{Kirsten}, F. and {Ould-Boukattine}, O.~S. and {Herrmann}, W. and {Gawro{\'n}ski}, M.~P. and {Hessels}, J.~W.~T. and {Lu}, W. and {Snelders}, M.~P. and {Chawla}, P. and {Yang}, J. and {Blaauw}, R. and {Nimmo}, K. and {Puchalska}, W. and {Wolak}, P. and {van Ruiten}, R.},
        title = "{A link between repeating and non-repeating fast radio bursts through their energy distributions}",
      journal = {Nature Astronomy},
         year = 2024,
        month = mar,
       volume = {8},
        pages = {337-346},
          doi = {10.1038/s41550-023-02153-z},
archivePrefix = {arXiv},
       eprint = {2306.15505},
 primaryClass = {astro-ph.HE},
       adsurl = {https://ui.adsabs.harvard.edu/abs/2024NatAs...8..337K}
}

@ARTICLE{curtin2024,
       author = {{Curtin}, Alice P. and {Sand}, Ketan R. and {Pleunis}, Ziggy and {Jain}, Naman and {Kaspi}, Victoria and {Michilli}, Daniele and {Fonseca}, Emmanuel and {Shin}, Kaitlyn and {Nimmo}, Kenzie and {Brar}, Charanjot and {Dong}, Fengqiu Adam and {Eadie}, Gwendolyn M. and {Gaensler}, B.~M. and {Herrera-Martin}, Antonio and {Ibik}, Adaeze L. and {Joseph}, Ronny C. and {Kaczmarek}, Jane and {Leung}, Calvin and {Main}, Robert and {Masui}, Kiyoshi W. and {McKinven}, Ryan and {Mena-Parra}, Juan and {Ng}, Cherry and {Pandhi}, Ayush and {Pearlman}, Aaron B. and {Rafiei-Ravandi}, Masoud and {Sammons}, Mawson W. and {Scholz}, Paul and {Smith}, Kendrick and {Stairs}, Ingrid},
        title = "{Morphology of 35 Repeating Fast Radio Burst Sources at Microsecond Time Scales with CHIME/FRB}",
      journal = {arXiv e-prints},
         year = 2024,
        month = nov,
          eid = {arXiv:2411.02870},
        pages = {arXiv:2411.02870},
          doi = {10.48550/arXiv.2411.02870},
archivePrefix = {arXiv},
       eprint = {2411.02870},
 primaryClass = {astro-ph.HE},
       adsurl = {https://ui.adsabs.harvard.edu/abs/2024arXiv241102870C}
}

@ARTICLE{yamasaki2024,
       author = {{Yamasaki}, Shotaro and {Goto}, Tomotsugu and {Ling}, Chih-Teng and {Hashimoto}, Tetsuya},
        title = "{The true fraction of repeating fast radio bursts revealed through CHIME source count evolution}",
      journal = {\mnras},
         year = 2024,
        month = feb,
       volume = {527},
       number = {4},
        pages = {11158-11166},
          doi = {10.1093/mnras/stad3844},
archivePrefix = {arXiv},
       eprint = {2309.14337},
 primaryClass = {astro-ph.HE},
       adsurl = {https://ui.adsabs.harvard.edu/abs/2024MNRAS.52711158Y}
}

@ARTICLE{Beniamini2025,
       author = {{Beniamini}, Paz and {Kumar}, Pawan},
        title = "{Can Repeating and Nonrepeating Fast Radio Bursts Be Drawn from the Same Population?}",
      journal = {\apj},
         year = 2025,
        month = nov,
       volume = {993},
       number = {1},
          eid = {37},
        pages = {37},
          doi = {10.3847/1538-4357/ae0712},
archivePrefix = {arXiv},
       eprint = {2506.09138},
 primaryClass = {astro-ph.HE},
       adsurl = {https://ui.adsabs.harvard.edu/abs/2025ApJ...993...37B}
}

@ARTICLE{nimmo2025,
       author = {{Nimmo}, Kenzie and {Pleunis}, Ziggy and {Beniamini}, Paz and {Kumar}, Pawan and {Lanman}, Adam E. and {Li}, D.~Z. and {Main}, Robert and {Sammons}, Mawson W. and {Andrew}, Shion and {Bhardwaj}, Mohit and {Chatterjee}, Shami and {Curtin}, Alice P. and {Fonseca}, Emmanuel and {Gaensler}, B.~M. and {Joseph}, Ronniy C. and {Kader}, Zarif and {Kaspi}, Victoria M. and {Lazda}, Mattias and {Leung}, Calvin and {Masui}, Kiyoshi W. and {Mckinven}, Ryan and {Michilli}, Daniele and {Pandhi}, Ayush and {Pearlman}, Aaron B. and {Rafiei-Ravandi}, Masoud and {Sand}, Ketan R. and {Shin}, Kaitlyn and {Smith}, Kendrick and {Stairs}, Ingrid H.},
        title = "{Magnetospheric origin of a fast radio burst constrained using scintillation}",
      journal = {\nat},
         year = 2025,
        month = jan,
       volume = {637},
       number = {8044},
        pages = {48-51},
          doi = {10.1038/s41586-024-08297-w},
archivePrefix = {arXiv},
       eprint = {2406.11053},
 primaryClass = {astro-ph.HE},
       adsurl = {https://ui.adsabs.harvard.edu/abs/2025Natur.637...48N}
}

@ARTICLE{eftekhari2025,
       author = {{Eftekhari}, T. and {Dong}, Y. and {Fong}, W. and {Shah}, V. and {Simha}, S. and {Andersen}, B.~C. and {Andrew}, S. and {Bhardwaj}, M. and {Cassanelli}, T. and {Chatterjee}, S. and {Coulter}, D.~A. and {Fonseca}, E. and {Gaensler}, B.~M. and {Gordon}, A.~C. and {Hessels}, J.~W.~T. and {Ibik}, A.~L. and {Joseph}, R.~C. and {Kahinga}, L.~A. and {Kaspi}, V. and {Kharel}, B. and {Kilpatrick}, C.~D. and {Lanman}, A.~E. and {Lazda}, M. and {Leung}, C. and {Liu}, C. and {Mas-Ribas}, L. and {Masui}, K.~W. and {Mckinven}, R. and {Mena-Parra}, J. and {Miller}, A.~A. and {Nimmo}, K. and {Pandhi}, A. and {Patil}, S.~S. and {Pearlman}, A.~B. and {Pleunis}, Z. and {Prochaska}, J.~X. and {Rafiei-Ravandi}, M. and {Sammons}, M. and {Scholz}, P. and {Shin}, K. and {Smith}, K. and {Stairs}, I.},
        title = "{The Massive and Quiescent Elliptical Host Galaxy of the Repeating Fast Radio Burst FRB 20240209A}",
      journal = {\apjl},
         year = 2025,
        month = feb,
       volume = {979},
       number = {2},
          eid = {L22},
        pages = {L22},
          doi = {10.3847/2041-8213/ad9de2},
archivePrefix = {arXiv},
       eprint = {2410.23336},
 primaryClass = {astro-ph.HE},
       adsurl = {https://ui.adsabs.harvard.edu/abs/2025ApJ...979L..22E}
}

@ARTICLE{sridhar2021,
       author = {{Sridhar}, Navin and {Metzger}, Brian D. and {Beniamini}, Paz and {Margalit}, Ben and {Renzo}, Mathieu and {Sironi}, Lorenzo and {Kovlakas}, Konstantinos},
        title = "{Periodic Fast Radio Bursts from Luminous X-ray Binaries}",
      journal = {\apj},
         year = 2021,
        month = aug,
       volume = {917},
       number = {1},
          eid = {13},
        pages = {13},
          doi = {10.3847/1538-4357/ac0140},
archivePrefix = {arXiv},
       eprint = {2102.06138},
 primaryClass = {astro-ph.HE},
       adsurl = {https://ui.adsabs.harvard.edu/abs/2021ApJ...917...13S}
}




\appendix

\section{UMAP hyperparameters}
\label{sec:appendixA}
In semi-supervised settings, the unlabeled (-1) samples can mix with true classes and contaminate the feature space. Rather than selecting parameters based solely on the highest clustering score, which can be sensitive to noise, contamination, and local irregularities, we instead adopt hyperparameters that fall within the standard stability regime and then assess their stability and representativeness after removing the class -1 samples. We adopt \texttt{n\_neighbors} $= 15$, which is consistent with the UMAP authors’ recommendation\footnote{\url{https://github.com/lmcinnes/umap}} that values between 5--50 (typically 10--15) provide a stable balance between local and global structure. This choice is further supported by applications across multiple domains in which \texttt{n\_neighbors} is often set to approximately 0.5--6\% of the dataset size \citep[e.g.][]{Heiser2020, Brown2023}. A local scan of \texttt{n\_neighbors} $\in [10,30]$ and \texttt{min\_dist} $\in [0,1]$ suggests silhouette scores\footnote{The silhouette score measures how similar a sample is to its own cluster compared with other clusters. Values range from -1 to 1, with higher scores indicating better-defined, well-separated clusters.} of $\sim 0.2$--0.3 (mean $\sim 0.22$), with our selected configuration (\texttt{n\_neighbors} $= 15$ and \texttt{min\_dist} $=0.1$) achieving 0.24 and ranking within the top 15\% of Calinski–Harabasz values\footnote{The Calinski–Harabasz, or variance ratio criterion, evaluates clustering quality by comparing between-cluster separation to within-cluster compactness; higher values indicate more distinct and well-defined clusters.} 

\begin{figure}
    \centering
    \includegraphics[width=20pc]{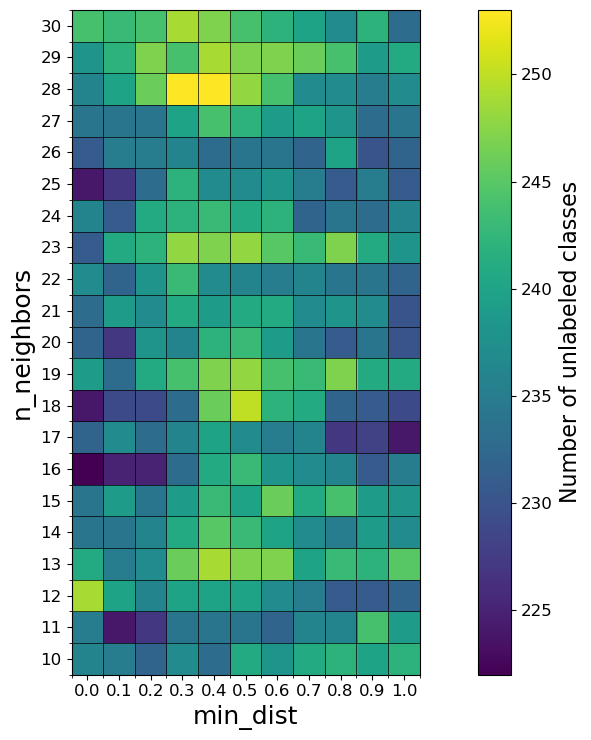}\\
    \caption{The heat map shows the number of unlabeled classes obtained across different hyperparameter settings. These hyperparameter pairs yield comparable silhouette scores of $\sim 0.2$--0.3, which is typical for UMAP embeddings with irregular or overlapping cluster shapes. Our selected hyperparameters yield 239 unlabeled sources, approximately the median of the values obtained across the search, with alternative choices differing by only $\sim 5$\% relative to our selected parameters. This small variation confirms the stability of our parameter choice.}
    \label{fig:grid_un}
\end{figure}

Fig.~\ref{fig:grid_un} shows how the number of samples assigned to the unlabeled class varies across different hyperparameter choices. Our selected hyperparameters yield 239 unlabeled sources, which is close to the median of the values obtained across the search. Furthermore, all tested hyperparameter pairs yield comparable scores, and the alternative choices within this search change the number of relabeled samples by only $\sim 5$\% relative to our selected configuration. This narrow variation demonstrates that our selected parameters fall within a stable region of the hyperparameter space, providing a representative and robust balance between local structure preservation and overall embedding consistency.

\section{Threshold Sensitivity Analysis}
\label{sec:appendixB}
We tested alternative neighborhood-fraction thresholds of 10\% and 30\%, in addition to the 20\%. The number of relabeled samples varies systematically with the threshold: 318, 239, and 152 for 10\%, 20\%, and 30\%, respectively. The corresponding numbers of recovered repeater candidates are 286, 168, and 117. This behavior is expected since higher thresholds send fewer non-repeater–like sources to the unlabeled pool, leaving most Class 0 samples confidently labeled from the outset and naturally yielding fewer candidates after self-training. Nevertheless, we also find that the downstream classification performance is $\sim 5$\% lower for the 10\% and 30\% thresholds compared with 20\%, indicating that 20\% provides the most balanced and effective model in our case.

Also, we would like to emphasize that the unlabeled-class threshold should not be set too high: at 30\%, too few samples are treated as ambiguous, reducing the informative unlabeled data needed for effective self-training. Conversely, a very low threshold such as 10\% sends too many samples to the unlabeled pool, introducing noise and weakening class structure. Therefore, the 20\% threshold offers the best compromise between these extremes and also achieves the highest performance. Importantly, nearly all repeater candidates identified at higher thresholds are recovered at lower thresholds, demonstrating the robustness of our core results.


\bsp	
\label{lastpage}
\end{document}